%% file: main_arxiv.tex
\documentclass[aps,pra,twocolumn,superscriptaddress,lo
ngbibliography,giveninits=true]{revtex4-1}

\AtBeginDocument{%
    \newwrite\bibnotes
    \def\bibnotesext{Notes.bib}
    \immediate\openout\bibnotes=\jobname\bibnotesext
    \immediate\write\bibnotes{@CONTROL{REVTEX41Control}}
    \immediate\write\bibnotes{@CONTROL{%
    apsrev41Control,author="08",editor="1",pages="1",title="0",year="1"}}
     \if@filesw
     \immediate\write\@auxout{\string\citation{apsrev41Control}}%
    \fi
}%

\usepackage{amsmath}
\usepackage{hyperref}
\hypersetup{
	colorlinks=true,
	linkcolor=blue,
	filecolor=blue,
	citecolor = blue,      
	urlcolor=blue,
}

\usepackage{graphicx}
\usepackage{dcolumn}
\usepackage{bm}
\usepackage{natbib}
\usepackage[utf8]{inputenc}
\usepackage{units}
\usepackage{amsmath}
\usepackage{amssymb}
\usepackage{graphicx}
\usepackage{bm}
\usepackage{multirow,microtype,color,relsize,ulem}

\usepackage[utf8]{inputenc}
\usepackage[T1]{fontenc}
\usepackage{mathptmx}
\usepackage{amsfonts}
\usepackage{amsmath}

\begin{document}

\title{Collective light shifts of many longitudinal cavity modes induced by coupling to a cold-atom ensemble}
\author{Marin Ðujić} 
\author{Mateo Kruljac} 
\author{Lovre Kardum} 
\author{Neven Šantić} 
\author{Damir Aumiler} 
\affiliation{Institute of Physics, Centre for Advanced Laser Techniques (CALT), Bijeni\v{c}ka cesta 46, 10000, Zagreb, Croatia}
\author{Ivor Krešić}\email{ikresic@ifs.hr}
\affiliation{Institute of Physics, Centre for Advanced Laser Techniques (CALT), Bijeni\v{c}ka cesta 46, 10000, Zagreb, Croatia}
\affiliation{Institute for Theoretical Physics, Vienna University of Technology (TU Wien), Vienna, A–1040, Austria}
\author{Ticijana Ban} 
\affiliation{Institute of Physics, Centre for Advanced Laser Techniques (CALT), Bijeni\v{c}ka cesta 46, 10000, Zagreb, Croatia}

\date{\today}

\begin{abstract}
We experimentally study the interaction between a cold atom cloud and many longitudinal modes of a high quality Fabry-Perot cavity, by measuring signatures of collective light shifts in the cavity transmission spectrum of an optical frequency comb probe. Using a resonator coupled to more than $10^5$ intracavity atoms, we detect significant shifts of $\sim 100$ cavity modes simultaneously, which is a direct manifestation of physics beyond the hitherto explored regime of cavity-cold atom interaction with only single or few longitudinal modes at a time. For the cavity mode closest to the atomic resonance, we demonstrate a bistability in the transmission spectrum, arising due to a combined coupling of the cloud to an external pump laser and a cavity mode probed by the optical frequency comb. These results provide a first step toward deeper exploration of multifrequency cavity quantum electrodynamics, where ultrashort pulsed sources could be used for optical manipulation, cooling and entanglement of cold atoms in a resonator.

\end{abstract}
\maketitle

%


\section{Introduction} 
Optical resonators have been a mainstay of atomic physics for the last four decades. The considerable enhancement of light-atom interaction, provided by coupling to a high quality optical cavity, has led to unprecedented levels of quantum control, down to single atoms and photonic quanta \cite{kimble98,raimond01,reiserer15}. For an ensemble of cold atoms, the dispersive regime of this interaction, occurring when resonator frequency is detuned sufficiently far from the optical transitions that resonant photon absorption is negligible, has also proven crucial for inducing spectacular collective effects such as cavity cooling \cite{domokos02,chan03,xu16,hosseini17}, squeezing \cite{leroux10,cox16} and self-organization \cite{black03,voncube04,arnold12,muniz20}. However, up to now collective effects in the dispersive regime of cavity-cold atom interaction were experimentally studied only using single and few longitudinal \cite{chan03,xu16,hosseini17,leroux10,cox16,black03,voncube04,arnold12,muniz20,guo2021optical,luo25} and many nearly-degenerate transverse \cite{kollar17,guo2021optical,kroeze_directly_2025,marsh25} cavity modes with continuous-wave (cw) laser driving. 

\begin{figure}[!h]
\centering
\includegraphics[clip,width=\columnwidth]{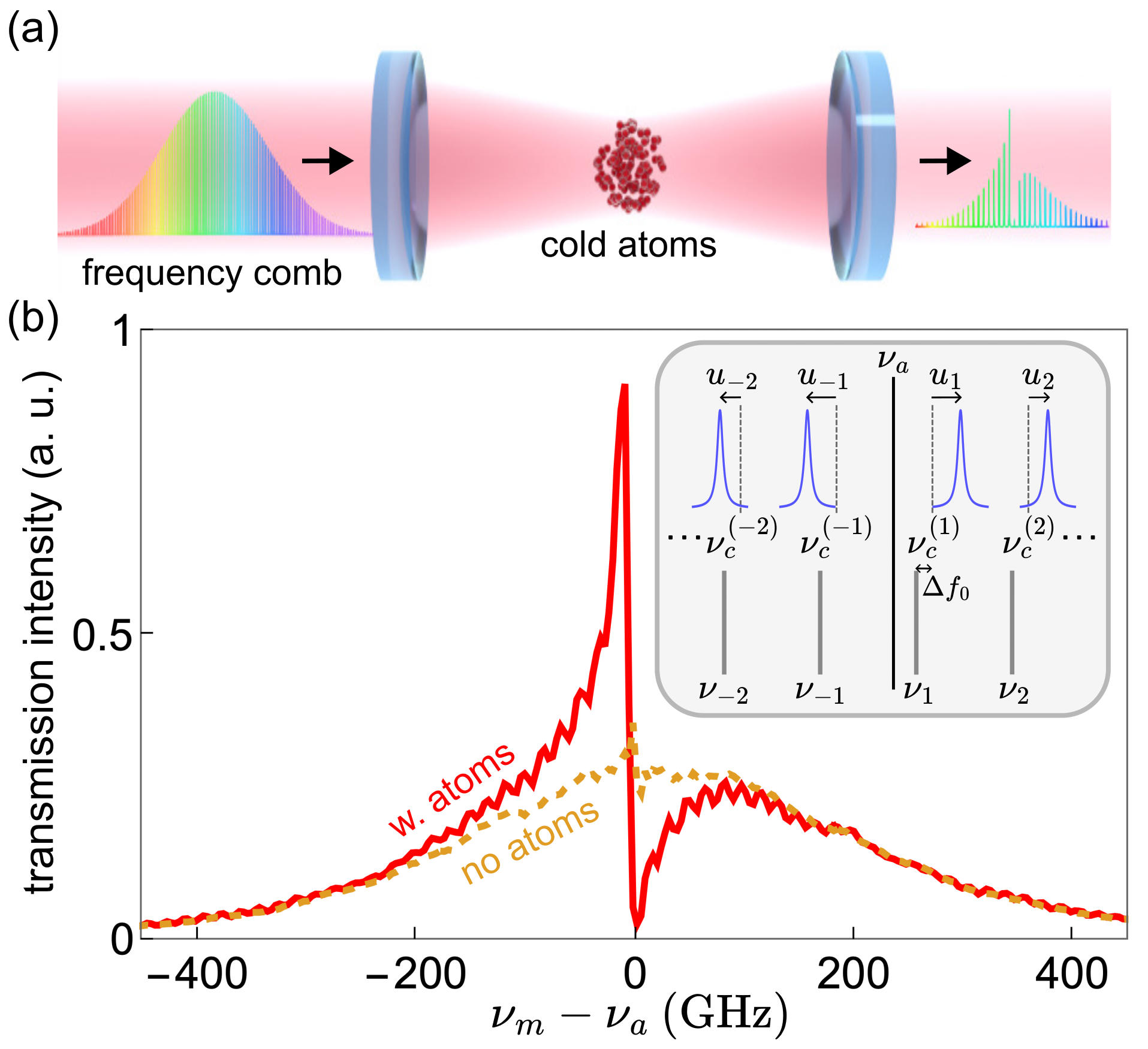}
\caption{Detecting collective light shifts of many longitudinal cavity modes induced by coupling to a cold atom ensemble. (a) Illustration of the experimental setup. Cavity transmission of an optical frequency comb (OFC) beam is measured using an optical spectrum analyzer. (b) Measurement results for an empty cavity (gold dashed line) and with atoms present (red solid line), under the condition of $\Delta f_0=-200$ kHz. Inset: $\nu_m$ - comb line frequencies, $\nu_c^{(m)}$ - bare cavity mode frequencies, $\nu_a$ - atomic transition frequency, $u_m$ - collective cavity light shifts induced by coupling to a cold atom ensemble, $\Delta f_0=\nu_1-\nu_c^{(1)}$.} 
\label{Fig:1}
\end{figure}

In contrast, recent theoretical studies highlight the exceptional promise of longitudinally multimode optical cavities with multifrequency drives, e.g. an optical frequency comb (OFC), for advancing cavity quantum electrodynamics (cQED). For instance, using an OFC excitation of many cavity modes can create a complex and highly controllable optical potential inside the cavity, which can be precisely shaped by tuning the OFC parameters, enabling the realization of effective atom-atom potentials beyond the all-to-all coupling paradigm \cite{ritsch13,mivehvar_cavity_2021}. An OFC-based approach has been proposed in \cite{masalaeva23} as a possible multimode cavity platform which can realize supersolids and droplets with phonon-like excitations. Multifrequency excitation with an OFC has also been suggested for implementing quantum annealing and Hopfield associative networks \cite{torggler17}, where for a $K$-site lattice the full site-to-site interaction matrix can be dynamically controlled by shaping up to $K(K+1)/2$ lasers derived from an OFC. Furthermore, driving a cold atom cloud with multifrequency laser light can lead to spontaneous formation of clusters in atomic density, the size of which is a direct signature of the drive and free spectral range of the cavity \cite{torggler20}. These numerical results also show that multifrequency driving can enhance cavity cooling compared to a single cw drive, which could enable cooling of atomic and molecular species with transitions at wavelengths below the optical region, where only ultrafast radiation sources are available \cite{kielpinski06}. Finally, a recent theoretical work has suggested that driving with weak optical pulses containing multiple spectral lines coinciding with cavity resonances can be used to project an atomic ensemble into an entangled state, which could allow heralded generation of complex nonclassical atomic states by detecting a single photon \cite{chen15}. 

A basic ingredient of all these proposals is longitudinally multimode cavity-atom coupling and multifrequency driving of the cavity, which was never before demonstrated for many modes in a cold-atom setup. We have now realized a platform exhibiting such features, demonstrating appreciable dispersive coupling between atoms and approximately 100 longitudinal cavity modes by measuring the transmission of an OFC through the cavity. Moreover, for the cavity mode nearest to the optical transition, we observe that the atomic saturation due to the presence of an external cooling laser driving the atoms leads to an optical bistability in the transmitted comb line, which extends the OFC driven cavity-atom interaction into the nonlinear regime.

\section{Cavity-transmitted comb spectra}
In the first set of experiments we detected the signatures of collective light shifts of many longitudinal cavity modes by measuring the transmission spectra of the OFC through the cavity-atom system. The basic principle of the experiment is schematically depicted in Fig. \ref{Fig:1}a). An OFC, with each 24th comb line tuned near a longitudinal (i.e. TEM$_{00}$) resonance of an empty cavity, is coupled into a two-mirror resonator with cold atoms loaded at its center. We denote each 24th OFC line's frequency as $\nu_m$ [see inset of Fig. \ref{Fig:1}b)], where e.g. $m=1$ ($m=-1$) is the index of the nearest blue (red) detuned line from the $^{87}$Rb D2-line $F=2\to F'=3$ atomic optical transition $\nu_a$. Analogously, the empty cavity mode frequency near the corresponding OFC line is denoted as $\nu_c^{(m)}$, and its collective light shift (see below) is denoted as $u_m$. Another important parameter is the detuning $\Delta f_0=\nu_1-\nu_c^{(1)}$, which can be precisely tuned in our setup (see below).

The OFC radiation transmitted through the cavity is collected onto an optical spectrum analyzer (OSA). The transmission spectra of the OFC beam at fixed OFC-cavity detunings $\Delta f_0$ contain information about the resonances in the atom-cavity system. By comparing to the empty cavity transmission and the theoretical results, the influence of atom-induced collective light shifts on many longitudinal cavity modes can then be inferred using a single OFC transmission measurement. The number of cavity modes shifted due to the presence of the atomic ensemble can thus be used to estimate how many longitudinal modes significantly participate in the atom-cavity coupling, which is a quantity of interest with regards to the theoretical proposals of \cite{masalaeva23,torggler17,torggler20,chen15,kielpinski06}.

\subsection{Experimental setup}
We now describe our experimental setup, with the main parameters for the cavity and OFC given in Table \ref{tab:params} and more details given in \cite{dujic25}. The cavity with length $L=7.76$ cm, mirror reflectivities $r=99.98\%$ and dispersion $\epsilon=18$ Hz (see Supplemental Materials for details \cite{dujic25}) is set up in the near-confocal configuration and stabilized via the Pound-Drever-Hall method. Every 24th OFC mode with frequency $\nu_m$ is matched to a cavity mode of frequency $\nu_c^{(m)}$, where the OFC is spatially matched to the TEM$_{00}$ cavity mode profile. The power per comb mode at the cavity input is $\approx 0.26$ $\mu$W. A power enhancement factor of up to $G\sim 400$ is reachable in our setup, resulting in a cavity-enhanced peak intensity of 332 mW/cm$^2$ per comb, where the mode waist radius is $w_0=100\:\mu $m. The OFC beam entering the cavity is circularly polarized. As the OFC frequency stabilization operates independently from the cavity stabilization, the OFC light can be freely switched on or off, and frequency detuned by a specified value $\Delta f_0$ relative to the cavity modes,  using an acousto-optic modulator (AOM).

The light transmitted through the cavity was collected onto an OSA with frequency resolution of $7.5$ GHz, such that individual transmitted OFC lines are not spectrally resolved. Note here that the bare cavity losses reduce the transmitted OFC power by a factor of $\sim 8$ with respect to the input. Also, the noise visible in the measured cavity transmission intensities in Figs. \ref{Fig:1}-\ref{Fig:3} was due to aliasing between the FSR and the resolution of the OSA. 

The number of longitudinal cavity modes significantly shifted due to collective light shifts for $\Delta f_0=-200$ kHz can be used to estimate the number of modes interacting appreciably strongly with the atomic ensemble, and was calculated from our measurements as follows. As shown in Fig. \ref{Fig:1}, the presence of atoms loaded into the center of the cavity induces frequency shifts of the cavity modes, resulting in a modification of the transmitted spectrum. Subtracting the transmission spectra measured with and without atoms reveals a differential signal that is nonzero in the frequency regions where mode shifts lead to changes in transmission, see Fig. \ref{Fig:2}c,d). In our analysis, a threshold of 5$\%$ of the maximum transmitted signal in the presence of atoms is used to define a deviation of the differential signal from zero. The number of affected cavity modes is then obtained by dividing the corresponding frequency interval by the FSR.

\begin{table}[t]
\begin{ruledtabular}
\begin{tabular}{lcccc}
 & FSR [GHz]   & $F$&$g_0/2\pi$ [kHz] & $\kappa/2\pi$ [kHz] \\
\hline
Cavity & 1.93  & $1.2\times 10^4$  & 140 & 150  \\ 
\hline 
\hline
\\[-9pt]
 & Bandwidth [THz] & $f_{rep}$ [MHz] & $\lambda_{center}$ [nm] & $N_{modes}$ \\
\hline
OFC & $\sim 2.5$ & 80.5 & 780 & $\sim 3\times 10^4$ \\
\end{tabular}
\end{ruledtabular}
\caption{\label{tab:params}%
Experimental parameter values for the Fabry-Perot cavity (FSR - free spectral range, $F$ - finesse, $g_0$ - cavity-atom coupling strength $\kappa $ - decay rate) and the optical frequency comb ($f_{rep}$ - repetition frequency, $\lambda_{center}$ - central wavelength, $N_{modes}$ - number of modes).}
\end{table}

The atoms are loaded into a magneto-optical trap (MOT) positioned to overlap with the center of the TEM$_{00}$ cavity mode. The MOT produces a cloud of $\approx 7 \cdot 10^6$ atoms at a temperature of around 70 $\mu$K and a radius of $\approx 0.55$ mm. As the repumper beam was turned on during these experiments, nearly all population was in the $F=2$ ground state. Since the $F=2\to F'=3$ transition is dominant in the $^{87}$Rb D2 line, we here treat the cloud as a two-level medium.

\subsection{Theoretical model}\label{sec:th_model_lin}
The atom-cavity behavior can be described using cQED \cite{tavis68}. The Hamiltonian modeling an OFC-driven longitudinal cavity modes coupled to $N$ static atoms placed at the cavity mode intensity maxima can be written as (see \cite{dujic25} for details of the derivations in this subsection):
\begin{align}\label{eq:h1at_3}
\begin{aligned}
\hat{H}(t) &=-\sum_{m}\Delta_{c,m}\hat{a}_{m}^\dagger\hat{a}_{m}+g_{0}\left[\sum_{m}\hat{a}_me^{-i\delta_{p,m}t}\hat{\sigma}_{eg}+\mbox{H.c.}\right]\\
+&i\sum_{m}\eta_m(\hat{a}_m^\dagger -\hat{a}_m ).
\end{aligned}
\end{align}
Here, $m$ is the cavity mode index with the summation going over all relevant longitudinal modes, $\hat{a}_m$ is the cavity destruction operator, $\eta_m\approx\eta$ is the driving rate (see below for the justification of the approximation), $\Delta_{c,m}=2\pi\times(\nu_m-\nu_c^{(m)})$ is the cavity-comb line detuning, $\Delta_a^{(m)}=2\pi\times(\nu_c^{(m)}-\nu_a)$ is the cavity-atom detuning, $\delta_{p,m}=2\pi\times(\nu_m-\nu_a)$ is the comb line-atom detuning, $\hat{\sigma}_z=\hat{\sigma}_{ee}-\hat{\sigma}_{gg}$, $\hat{\sigma}_{ge}$ are the collective atomic optical inversion and coherence operators for the $N$ atoms and $g_0$ is the cavity-atom coupling strength.

The effective Hamiltonian of (\ref{eq:h1at_3}) in the dispersive limit can be calculated using the Schrieffer-Wolff transformation \cite{schrieffer66} while keeping the terms $\propto  g_0^2/\Delta_{a}^{(m)},\: g_0\eta_m/\Delta_{a}^{(m)}$ and neglecting the cavity mediated dipole-dipole interaction \cite{lehmberg70} and cavity mode-mode scattering \cite{wickenbrock_collective_2013}. This Hamiltonian is given by: 
\begin{align}\label{eq:Hamiltonian}
\begin{aligned}
&\hat H_{eff}(t)= -\sum_m \Delta_{c,m}\,\hat a_m^\dagger\hat a_m
   + i\sum_m \eta(\hat a_m^\dagger-\hat a_m) \\
   -&\sum_m \frac{g_0^2}{\Delta_{a}^{(m)}}\left(\hat a_m^\dagger\hat a_m+\frac{1}{2}\right)\hat\sigma_z+ \sum_m \left(\frac{ig_0\eta}{\Delta_{a}^{(m)}}e^{i\delta_{p,m}t}\hat\sigma_{ge}+\mbox{h.c.}\right).
\end{aligned}
\end{align}
The first two terms in $\hat{H}_{eff}(t)$ describe the cavity modes and OFC driving. The third term describes the collective light shifts of the cavity modes and the light shift of the atomic transition, arising due to atom-cavity coupling. The last term describes the effective optical driving of the atomic transition due to the cavity modes pumped by the OFC, similarly to free-space case \cite{felinto03,marian04,marian05,jayich16,santic19}.

Using $\hat{H}_{eff}(t)$ for calculating the steady state mean field cavity mode population $|\alpha_m|^2=\langle\hat{a}_m^\dagger\hat{a}_{m}\rangle/N$ in the thermodynamic linear optics limit leads to:
\begin{align}\label{eq:cavity}
\begin{aligned}
|\alpha_m|^2 &= \frac{|\eta'|^2}{(\Delta_{c,m}-U_m)^2+\frac{\kappa^2}{4}},
\end{aligned}
\end{align}
where $\eta'=\eta/\sqrt{N}$ and $\kappa$ is the cavity decay rate. The $m$-th cavity mode thus experiences a collective light shift with respect to the empty cavity given by $U_m=2\pi\times(\nu_{c,m}'-\nu_{c}^{(m)})=2\pi\times u_m=Ng_0^2/\Delta_a^{(m)}$. We note that although the shifts were here derived using cQED, their classical interpretation is the refractive index of the cold atomic medium changing the optical length of the resonator and altering the cavity resonance condition \cite{dujic25}.

\subsection{Experimental results}
For an empty cavity [see gold dashed line in Fig. \ref{Fig:1}b)], each of the narrow OFC lines with frequency near a cavity resonance follows the characteristic Lorentzian transmission curve of linewidth $\kappa/2$, while other OFC modes are reflected. An important quantity here is $\Delta f_0$, defined by the difference of the first OFC mode from its nearest cavity mode, see inset of Fig. \ref{Fig:1}b). At some fixed value of $\Delta f_0$, the OFC transmission through the $m$-th cavity resonance does not equal the transmission through the first cavity resonance, due to cavity dispersion $\epsilon$ caused by the mirror dispersion \cite{dujic25,thorpe05}. Instead, the detuning of the $m$-th OFC line's frequency $\nu_m$ from its nearest empty cavity mode $\nu_{c}^{(m)}$ is given by $\nu_m-\nu_{c}^{(m)}=\Delta f_0-|m|(|m|-1)\epsilon/2$. This means that the transmitted signal will have a bandwidth significantly smaller than the input OFC. This justifies the approximation $\eta_m\approx \eta$ in Eq. (\ref{eq:Hamiltonian}), since the input OFC spectrum of bandwidth $2.5$ THz varies negligibly on the cavity transmission spectrum of bandwidth $\sim 400$ GHz. It is worth noting that in our experiment the OFC spectrum can be frequency-shifted by an AOM before being coupled into the cavity (see \cite{dujic25}). This enables precise matching of a selected comb mode to the corresponding cavity resonance, e.g. $\nu_1$ to mode $\nu_{c}^{(1)}$, as well as controlled scanning of the comb mode around that resonance.

\begin{figure}[!t]
\centering
\includegraphics[clip,width=\columnwidth]{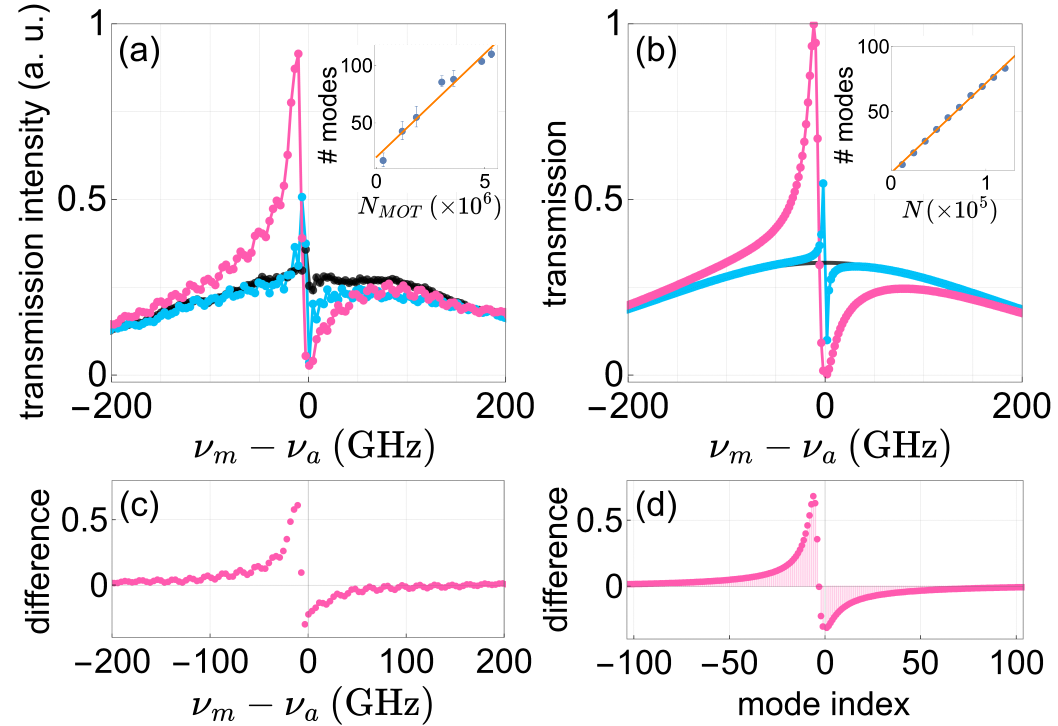}
\caption{Atom number dependence of the OFC cavity transmission spectrum. (a) Experimental and (b) simulation data for empty cavity (black), partially loaded MOT (blue) and fully loaded MOT (magenta). Lines are guide to the eyes. Insets: Atom number scans of (a) experimental and (b) theoretical number of cavity modes significantly shifted due to the presence of atoms (see text). (c) Experimental and (d) simulated difference in the transmission value with respect to empty cavity for the highest number of atoms. Experimental parameters: MOT atom number $N_{MOT}=(0,1.2,5.3)\times10^6$ for black, blue and pink, $\Delta f_0=-200$ kHz. Simulation parameters: $\mbox{FSR}=1.93$ GHz, $g_0=2\pi\times 140$ kHz, $\Delta_a^{(1)}=2\pi\times 495$ MHz, $\epsilon=18$ Hz, mirror transmission coefficient $t=0.0125$, $N=(0,0.06,\:1.2)\times 10^5$ for black, blue and magenta and $\Delta f_0=-220$ kHz.}
\label{Fig:2}
\end{figure}  

Deriving the Lorentzian profile (\ref{eq:cavity}) and shift $u_m$ for each longitudinal mode with index $m$ allows now to explain the spectral shape of the OFC transmission through the cavity, shown for a particularly notable case of $\Delta f_0=-200$ kHz in Fig. \ref{Fig:1}b). For cavity modes to the red of the atomic transition ($\Delta_a^{(m)}<0$), the collective shift will reduce the cavity frequencies, and conversely increase them for the blue detuned cavity modes, see inset of Fig. \ref{Fig:1}b). For this $\Delta f_0$, the small $|\Delta_a^{(m)}|$ modes with $\Delta_a^{(m)}<0$ shift the cavity resonances closer to the OFC lines, resulting in a strong increase in transmission. In contrast, for the small $|\Delta_a^{(m)}|$ modes with $\Delta_a^{(m)}>0$, the transmission is suppressed nearly to zero due to collective light shifts moving the cavity modes away from OFC lines. As detailed below, this constitutes clear signatures of simultaneous dispersive cavity-atom interaction with many longitudinal cavity modes.

We now discuss the experimental results of cavity transmission spectrum measurements for $\Delta f_0=-200,\:0,\:200$ kHz and give comparisons to the theoretical predictions for the total OFC transmission signal, calculated by summing over the transmission profiles of each cavity mode [see Eq. (S25) of \cite{dujic25}]. We note that all of the experimental data for different $\Delta f_0$'s has been normalized to the same maximum. For $N=0$, the theoretical curves differ from the experimental ones near $\nu_m-\nu_a=0$. This small central feature in the OFC transmission signal for a nominally empty cavity is a consequence of the residual background atomic vapour coupling to the cavity and shifting the cavity modes.

The $\Delta f_0=-200$ kHz results for different $N$ values are depicted in Fig. \ref{Fig:2}. When cold atoms are loaded into the cavity, a clear signature of the collective light shifts can be detected in the strong increase of cavity transmission for $m$ values in the approximate range $-50 <m \leq -4$. For modes in the range $-4<m<45$, the collective light shift pushes the cavity modes away from the OFC modes, which results in a reduction of cavity transmission, reaching values near zero. For $m$'s outside of these two ranges, the $|\Delta_a^{(m)}|$ increases and $|u_{m}|$ is reduced far below $\kappa/2\pi$, such that eventually the transmission spectrum resembles that of an empty cavity. The transmission curves are in good agreement with the theoretical calculations shown in Figs. \ref{Fig:2}b,d). The number of cavity modes that are simultaneously shifted due to the cavity-atom interaction grow proportionally to the number of atoms loaded into the cavity waist, see insets of Figs. \ref{Fig:2}a,b). From this data we conclude that appreciable simultaneous collective light shifts of more than 100 cavity modes are achieved in the setup for the highest number of atoms in the MOT. Note that here the MOT number $N_{MOT}$ and cavity atom number $N$ differ by a factor of $\gtrsim 10$. This is a consequence of both the atomic motion inside the cavity modes reducing the effective atom-cavity coupling $g_0$, and the non-optimal overlap of the cloud density distribution with radius $500\:\mu$m to the mode waist with radius $w_0=100\:\mu$m.

\begin{figure}[!t]
\centering
\includegraphics[clip,width=\columnwidth]{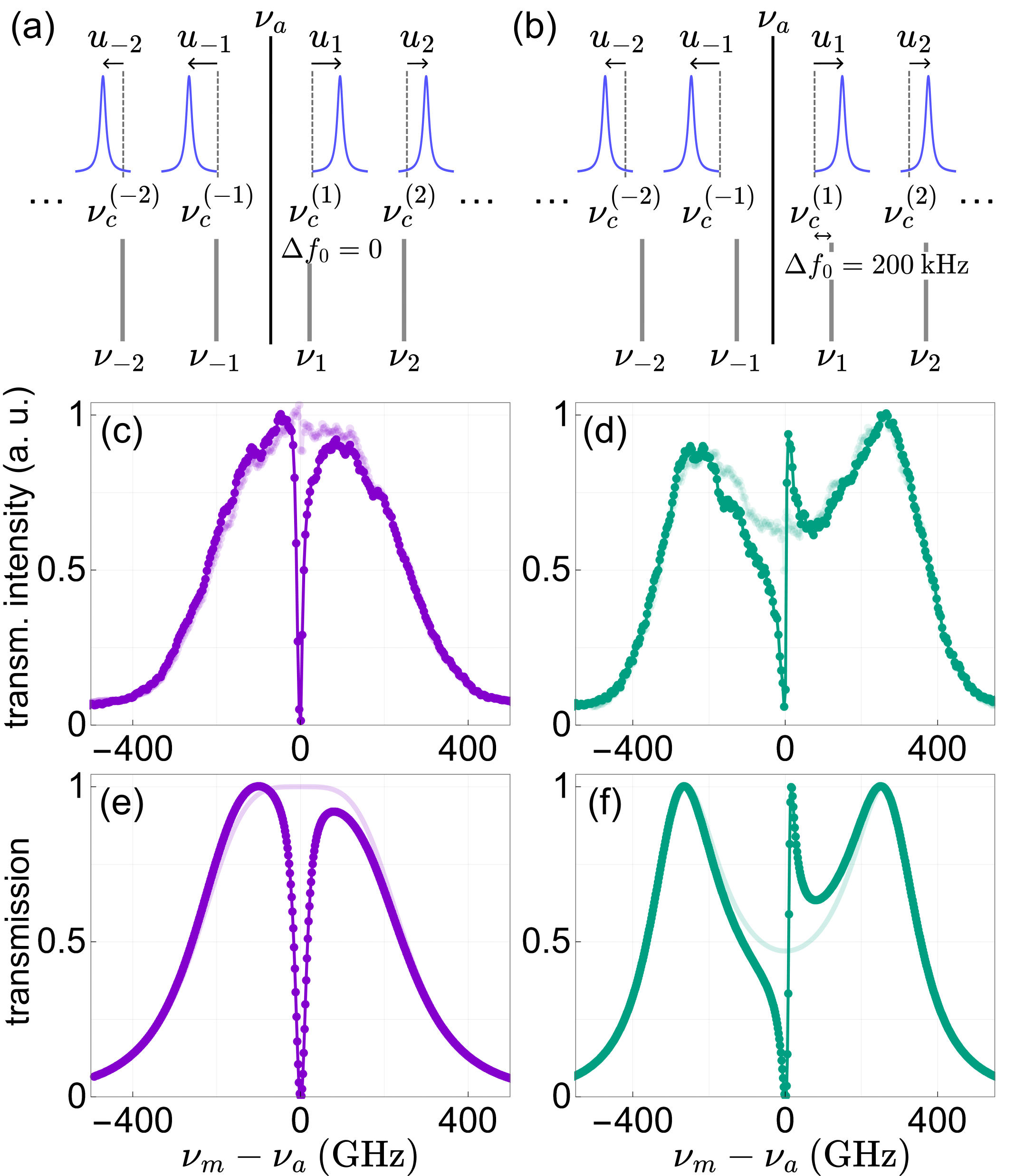}
\caption{Cavity transmission spectra for OFC tuned resonantly and to the blue of the empty cavity modes. Frequency schematics of the first several modes for (a) $\Delta f_0=0$ and (b) $\Delta f_0=200$ kHz. (c, d) Experimental and (e, f) simulation data for the resonant and blue-detuned cases, respectively. Thick lines - cavity with atoms loaded, semitransparent lines - no atoms in the cavity. Experimental parameters: see text. Simulation parameters: $\mbox{FSR}=1.93$ GHz, $g_0=2\pi\times 140$ kHz, $\Delta_a^{(1)}=2\pi\times 495$ MHz, $\epsilon=18$ Hz, mirror transmission coefficient $t=0.0125$, $N=1.2\times 10^5$ and $\Delta f_0=0,\: 160$ kHz for (e,f) respectively.}
\label{Fig:3}
\end{figure} 

The results for $\Delta f_0=0$ are given in Figs. \ref{Fig:3}a,c,e). For an empty cavity, the transmission peak here reaches a maximum value, theoretically equal to unity, near $\nu_m=\nu_a$. When atoms are loaded into the cavity, a transmission dip appears in the center of the spectrum, which is caused by the collective light shift pushing the cavity modes out of resonance with frequencies in the approximate range $-100\mbox{ GHz}<\nu_m-\nu_a<100\mbox{ GHz}$. The slight asymmetry in the left and right transmission peaks is a consequence of the asymmetry in $\Delta_a^{(1)}=2\pi\times 495\mbox{ MHz}=81.6\Gamma$ and $\Delta_a^{(-1)}=-2\pi\times1437\mbox{ MHz}=-236.6\Gamma$, which causes larger $|u_m|$ values to appear for modes with $m>0$.

The transmission for $\Delta f_0=200$ kHz is shown in  Fig. \ref{Fig:3}b,d,f). For an empty cavity, the transmission is small at frequencies near the center and there are two symmetric peaks at values $\nu_m-\nu_a\approx\pm 260$ GHz, where the cavity dispersion results in resonances for the OFC. A slight asymmetry in the peak heights in the experimental results is caused by a mismatch between the position of the center of the OFC emission spectrum and $\nu_a$. When atoms are loaded into the cavity, the cavity modes near the center shift towards higher frequencies for positive $m$, and towards lower frequencies for negative $m$ values. This causes a transmission increase for OFC modes with $m>0$ and a decrease for $m<0$, resulting in a curve with an inverted shape to the case of $\Delta f_0=-200$ kHz for small $|m|$ values. The theory reproduces the experimental data reasonably well, with the main discrepancies stemming from uncertainties in estimating experimental parameters such as the detuning of the relative OFC-cavity detuning $\Delta f_0$, cavity dispersion $\epsilon$ and reflectivity $r$, and cavity linewidth $\kappa$.

\section{Resolving individual cavity resonances}
\subsection{Experimental setup}
In the second set of experiments, we directly measured the transmission curves of cavity modes with $m=\pm 1,\pm 2$ in the single comb line coupling regime, where only one OFC mode is coupled to the cavity at a time, see \cite{dujic25} for details. The very narrow linewidths of individual comb lines ($\sim 10$ kHz) allow for transmission measurements which can resolve the shape of individual cavity modes. 

To achieve the regime of effectively single comb line cavity coupling, the empty cavity resonances were shifted by $u_{m}$ via loading the MOT, after which $\nu_m-\nu_c^{(m)} $ was scanned to find the new OFC-cavity resonance conditions for each $m = \pm 1, \pm 2 $ mode. Since $u_m$ for such small $m$ values exceeds $\kappa/2\pi $, shifting the OFC-cavity detuning $\nu_m-\nu_c^{(m)} $ by $ u_m $ tunes all other comb modes away from their corresponding cavity resonances, i.e. only a single comb mode is resonant with a cavity mode at a time, excluding the far-detuned background modes with very large $m$. The single-comb coupling condition was experimentally verified by measuring the cavity transmission spectrum using an OSA, which consisted of a narrow spectral feature limited by the resolution of the OSA, along with a wide background for large $|\nu_m-\nu_a|$ values. Due to this background signal, the OFC cavity transmission cannot be detected using the standard methods, and it was necessary to implement a heterodyne detection technique, similar to that in \cite{kresic19}, and described in detail in \cite{dujic25}, which enabled measurements of the cavity transmitted electric field of a single $m$-th comb line at different values of $\nu_m-\nu_c^{(m)} $. 

\subsection{Experimental results}
\begin{figure}[!t]
\centering
\includegraphics[clip,width=0.95\columnwidth]{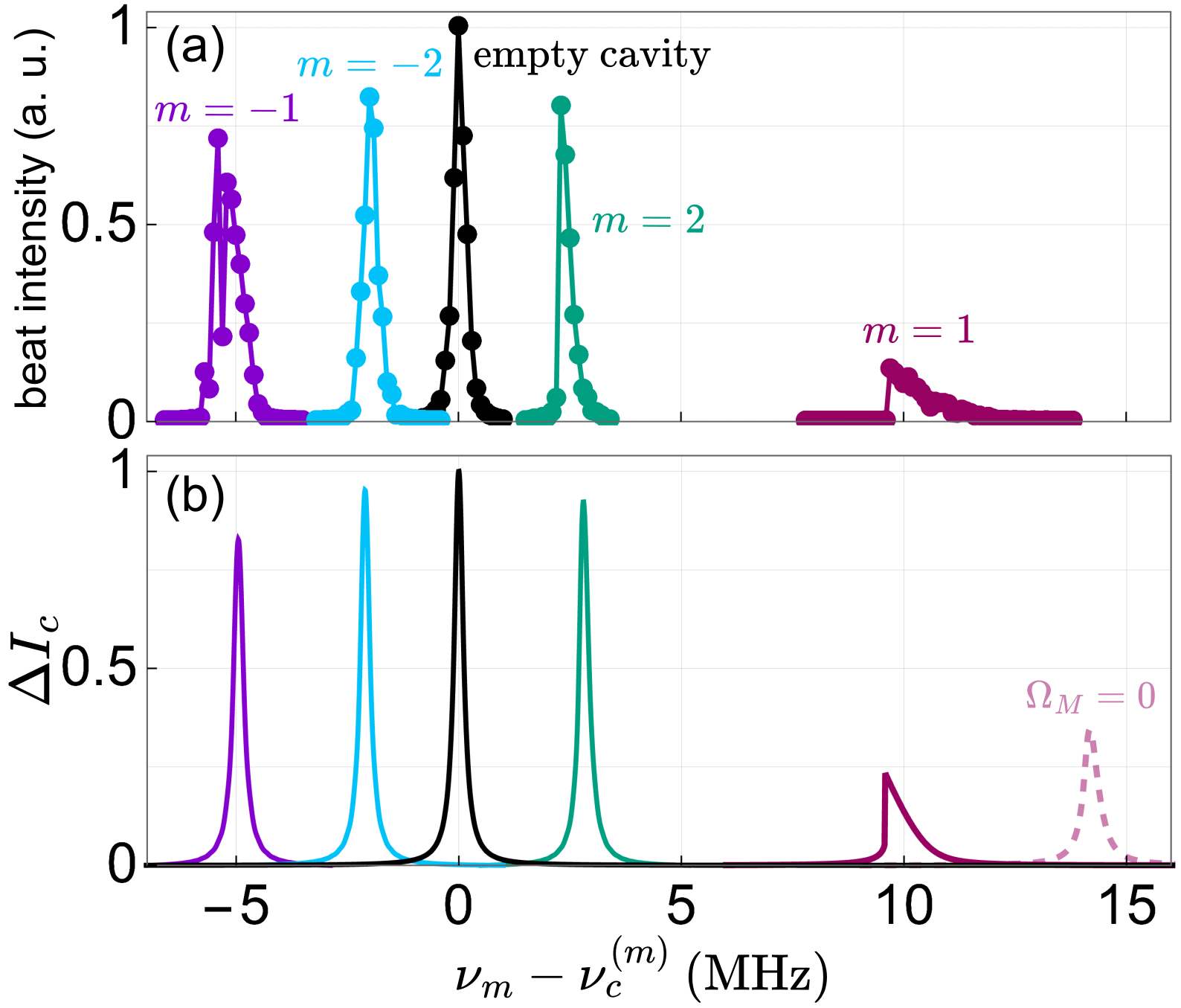}
\caption{Cavity transmission of individual comb lines in the single comb mode coupling regime. (a) Experimental measurements of transmission curves for modes with $m=1$ (red),  $m=-1$ (purple), $m=2$ (green), $m=-2$ (blue) and for an empty cavity mode (black). (b) Simulated difference in mean cavity photon number when the comb beam is turned on and off (see text), normalized to the empty cavity value. For comparison, the curve for the $m=1$ mode with $\Omega_M=0$ is shown. Experimental parameters: see text. Simulation parameters: $N=3.7\times 10^5$, $\kappa=2\pi\times 0.24$ MHz, $g_0=2\pi\times 140$ kHz, $\Omega_M=2\pi\times 1.7$ MHz, $\eta/\sqrt{N}=2\pi\times 0.06$ MHz, $\Gamma=2\pi\times 6.066$ MHz, $(\Delta_a^{(1)},\Delta_a^{(-1)},\Delta_a^{(2)},\Delta_a^{(-2)})=2\pi\times( 495,-1437,2427,-3369)$ MHz and $\Delta_M^{(m)}=\Delta_a^{(m)}+2\Gamma$.}
\label{Fig:4}
\end{figure}

The measurement results for the single comb mode coupling regime are given in Fig. \ref{Fig:4}a). Just as in the previous section, the presence of atoms shifts the central frequencies of the cavity modes by the collective light shifts given by $u_m=Ng_0^2
/(2\pi\Delta_a^{(m)})$. For modes with $m=-1,\pm 2$, the cavity resonance positions follow this analytic prediction derived from the linear optics approximation in Sec. \ref{sec:th_model_lin}. The transmission curves of the $m=\pm 2$ modes (and higher $|m|$ modes) follow a Lorentzian shape (\ref{eq:cavity}), also measured for an empty cavity. Note that the empty cavity linewidth is measured to be $\kappa/2\pi=240$ kHz, which differs from the value of $\kappa/2\pi=150$ kHz measured by cavity ringdown (see Table \ref{tab:params}), where the broadening is here induced by the cavity length stabilization. In contrast, for the $m=1$ mode, and to a lesser degree the $m=-1$ mode, there is a deviation from the Lorentzian cavity resonance shape. Particularly, the shape of the $m=1$ resonance is characteristic of bistable nonlinear cavities, for which multiple transmission values coexist at the same input frequency and intensity \cite{gibbs76,gothe19}. Note that this bistable transmission for $m=1$ should be present also in the results of Figs. \ref{Fig:1}-\ref{Fig:3}. However, a single mode influences the entire optical spectrum of the cavity-transmitted OFC beam sufficiently little that it can be neglected on the larger scale, such that the linear model reproduces the data well.

Describing the transmission shape of the $m=1$ cavity mode requires thus a treatment beyond the linear optics regime. The peak cavity intensity per comb mode is $\sim 330$ mW/cm$^2$, which results in a detuned saturation parameter of $s\sim 0.005$ for the far-detuned $\nu_1$ comb line with $I_s=2.5$ mW/cm$^2$, such that the population in the excited state is $\sigma_{ee}\approx s/2= 0.0025$. The bistable transmission curve in Fig. \ref{Fig:4}a) should thus not be a consequence of saturating the two-level transition by only the $\nu_1$ comb line. Magnetic sublevel pumping nonlinearities have been known to cause bistable transmission at low powers for a cavity coupled to a cold atom cloud \cite{lambrecht95}. However, the $^{87}$Rb D2-line $F=2\to F'=3$ transition has a self-defocusing magnetic pumping nonlinearity for blue detuning \cite{labeyrie25}, which would result in a cavity transmission curve with reversed shape compared to the one in Fig. \ref{Fig:4}a) for the $m=1$ mode. Additionally, reducing the repumper beam intensity did not change the shape of the $m=1$ transmission curve in our experiments, provided that the MOT atom number did not decay. This indicates that the optical nonlinearity is not a consequence of pumping the atoms into the $F=1$ state. 

We identify the main cause of the nonlinearity to be the presence of MOT cooling beams driving the $F=2\to F'=3$ transition, kept on during the measurements. This interpretation is corroborated by additional transient transmission measurements for a single comb line at a fixed frequency corresponding to the peak of the $\nu_c^{(1)}$ cavity mode transmission [see Fig. 4(a)]. In this experiment, the MOT laser beams were switched on and off using an AOM, and the resulting time dependence of the cavity transmission was recorded. We observed that the transmitted signal at the selected frequency appears only when the MOT beams are on. When the MOT beams are turned off, the cavity transmission rapidly vanishes, caused by a shift of the cavity mode to a frequency different from the one being probed [see also dashed line in Fig. \ref{Fig:4}b)]. The vanishing of cavity transmission occurs on timescales $\ll 1$ ms, such that the density of the cold atom cloud does not change appreciably. These preliminary measurements thus indicate that the cooling beam helps drive the saturation of the two-level transition and cause the bistable transmission curve, as corroborated by simulation results below. The measured cavity bistability will be explored in more detail in a subsequent publication.

\subsection{Numerical simulations}
As the cooling beams were kept on during the measurements, we here model the system using a cavity-atom interaction Hamiltonian of the form $\hat{H}_{int}(t)=(g_0\hat{a}+\Omega_{M}e^{i\Delta_{M}t})\hat{\sigma}_{eg}+\mbox{H.c.}$, where $\hat{\sigma}_{eg}$ is the collective atomic ``spin" raising operator and $\hat{a}$ is the cavity mode annihilation operator of the single cavity mode that is excited \cite{dujic25}. In the thermodynamic mean field limit, where $\langle\hat{O}_1\hat{O}_2\rangle\to\langle\hat{O}_1\rangle\langle\hat{O}_2\rangle$ and $\langle \hat{a}\rangle\to \sqrt{N}\alpha$, $\langle\hat{\sigma}_{ij}\rangle\to N\sigma_{ij}$, the equations of motion are given by: 
\begin{equation}\label{eq:MOT_comb}
\begin{aligned}
\frac{\partial \alpha}{\partial t}&=\left(i\Delta_{c}-\frac{\kappa}{2}\right)\alpha-ig_N\sigma_{ge}+\frac{\eta}{\sqrt{N}},\\
\frac{\partial \sigma_{ee}}{\partial t}&=-\Gamma\sigma_{ee}+\left[i\left(g_N\alpha^*+\Omega_M^*e^{-i\Delta_M t}\right) \sigma_{ge}+\mbox{H.c.}\right] ,\\
\frac{\partial \sigma_{eg}}{\partial t}&=-\left(i\Delta_{a}+\frac{\Gamma}{2}\right)\sigma_{eg}+i\left(g_N\alpha^*+\Omega_M^*e^{-i\Delta_M t}\right) (1-2\sigma_{ee}),
\end{aligned}
\end{equation}

where $g_N=\sqrt{N}g_0$. Here, a single comb line is driving the cavity mode while a second laser, detuned from the comb line by $\Delta_M$, is driving the atomic two-level transition directly with Rabi frequency $\Omega_M$. The above equations are solved numerically, and the quantity $\Delta I_c$, defined as the temporally averaged steady state difference between the cavity mode populations $|\alpha|^2=\langle\hat{a}^\dagger\hat{a}\rangle/N$ when the comb beam is turned on and off, is plotted in Fig. \ref{Fig:4}b). For $m=\pm 2$, the simulations reproduce the cavity mode shifts and shapes of the experiment, while for $m=-1$ the measured transmission line is broadened with respect to the simulated curve. The reason for this discrepancy is currently under investigation.

For the $m=1$ mode, the cavity photon number has a notably different shape and resonance position when turning on the cooling beam. At $\Omega_M=0$ (dashed curve), the cavity resonance has a Lorentzian shape. Increasing the value of $\Omega_M$ (solid curve) to $\Omega_M=2\pi\times 1.7$ MHz, the simulations reproduce the shape and position of the experimental transmission curve well. Indeed, in simulations with $\Omega_M=0$, the maximum $\sigma_{ee}$ reaches the experimentally estimated value of 0.0025 (see above), while for the finite $\Omega_M$ the value is $\sim 0.2$, which indicates the presence of the nonlinear regime in the two-level atomic response, as also experimentally expected from the cooling beam intensities (see below). We note here that the only free parameter used in the $m=1$ simulations is $\Omega_M$, while $\kappa$, $N$ and $\eta$ values were determined from the empty cavity transmission values, the $m=-1,\pm2$ transmission values, and the experimental saturation parameter, respectively, with $g_0,\Delta_a^{(m)},\Gamma$ taken directly from experiment.

Noting that the Rabi frequency of the MOT laser is given by $\Omega_M=\Gamma\sqrt{I_M/2I_s}$ \cite{foot_atomic_2005}, for saturation intensity $I_s=2.5$ mW/cm$^2$ the intensity for $\Omega_M=2\pi\times 1.7$ MHz is 0.4 mW/cm$^2$. This value is smaller than the total experimental cooling beam intensity of 20 mW/cm$^2$, which was taken from power measurements after an optical fiber. These measurements overestimate the total MOT beam intensity at cavity center as they did not include effects such as losses at optical elements and vacuum chamber windows, along with the absorption of the cooling beam within the cold cloud.

\section{Conclusion} 
We have experimentally demonstrated the interaction between cold atoms and $\sim 100$ longitudinal cavity modes by measuring the signatures of collective light shifts in the transmission of an incoming OFC beam. Additionally, we have measured a bistable transmission for the cavity mode nearest to the atomic transition, demonstrating nonlinear interaction due to combined coupling of the atoms to an external pump laser and the OFC driven cavity. 

By combining three experimental platforms hitherto never used together in the same setup: (i) a cold atom ensemble coupled to (ii) an optical resonator with (iii) ultrafast laser driving, our experiment has the potential to establish a novel avenue for studying light-matter interaction. The advantage offered by each platform: narrow atomic transition linewidths, interaction enhancement via the cavity, and multifrequency interfacing via the OFC, could enable a host of developments in the field, as envisioned by recent theoretical proposals towards utilizing this system for studying multifrequency cavity QED \cite{masalaeva23,torggler17,torggler20,chen15}.

The experiment can readily be extended into several intriguing directions. For instance, at mirror distances tuned close to the confocal regime, transverse cavity modes with frequency separations on $\sim 10$ MHz scales can be achieved \cite{kollar17}. Using an OFC pump with tailored $f_{rep}$ \cite{canella_low-repetition-rate_2024} to drive the atoms in such a configuration could lead to additional flexibility in designing complex effective cavity-mediated interaction Hamiltonians. Moreover, effects such as mode-mode scattering for nearly degenerate transverse cavity modes could be extended to the regime of OFC driving, potentially enabling multiplexed information processing at many cavity frequencies \cite{wickenbrock_collective_2013}. Also, utilizing two-photon atomic transitions in the multifrequency cavity context could lead to other interesting effects in analogy with the free-space configuration \cite{jayich16}. Regarding intensity squeezing of light with a nonlinear medium in a cavity \cite{lambrecht96}, using OFC driving with very high power per comb mode could here lead to frequency multimode squeezed light interfaced with atoms in the cavity - currently a highly sought-after goal in quantum photonics \cite{chen14, roslund14}.

Finally, we note that at present there is a great need for experimental techniques for manipulating atoms with transitions in the vacuum- and extreme-ultraviolet regions \cite{barry_cold_2016}, where only pulsed radiation sources are available \cite{gohle05,rodriguez2016}, along with molecules with complex transition level structures, where frequency combs could address multiple degrees of freedom with a single laser \cite{kielpinski06}. Exploration of cavity-cold atom interaction with OFC driving is a step towards developing new experimental methods applicable in such situations.

\textit{Acknowledgements.} We thank Helmut Ritsch and Sina Zeytinoglu for helpful discussions. Our work was supported by the project Transport, metastability, and neuromorphic applications in quantum networks (QNet), funded within the QuantERA II Programme that has received funding from the EU’s H2020 research and innovation programme under the GA No 101017733, and with funding organization Croatian Science Foundation (HRZZ).
We also acknowledge support from the project Centre for Advanced
Laser Techniques (CALT), cofunded by the European Union through the European Regional Development Fund under the Competitiveness and Cohesion Operational Programme (Grant No. KK.01.1.1.05.0001). IK acknowledges support from the project KODYN financed by the European Union through the National Recovery and Resilience Plan 2021–2026, and the Croatian Science Foundation (HRZZ) grant IP-2024-05-5670.

\bibliography{references}
\clearpage
\newpage
\renewcommand{\appendixname}{Supplemental Material}

\appendix
\input{sup_final.tex}


\end{document}

%% file: sup_final.tex
\title{Supplemental Material: Collective light shifts of many longitudinal cavity modes induced by coupling to a cold-atom ensemble}
\setcounter{equation}{0}
\setcounter{figure}{0}
\renewcommand{\theequation}{S\arabic{equation}}
\renewcommand{\theHsection}{S\arabic{section}}
\renewcommand{\thefigure}{S\arabic{figure}}

\maketitle


\section{Experimental setup and measuring techniques}

\begin{figure*}[!t]
\centering
\includegraphics[clip,width=2\columnwidth]{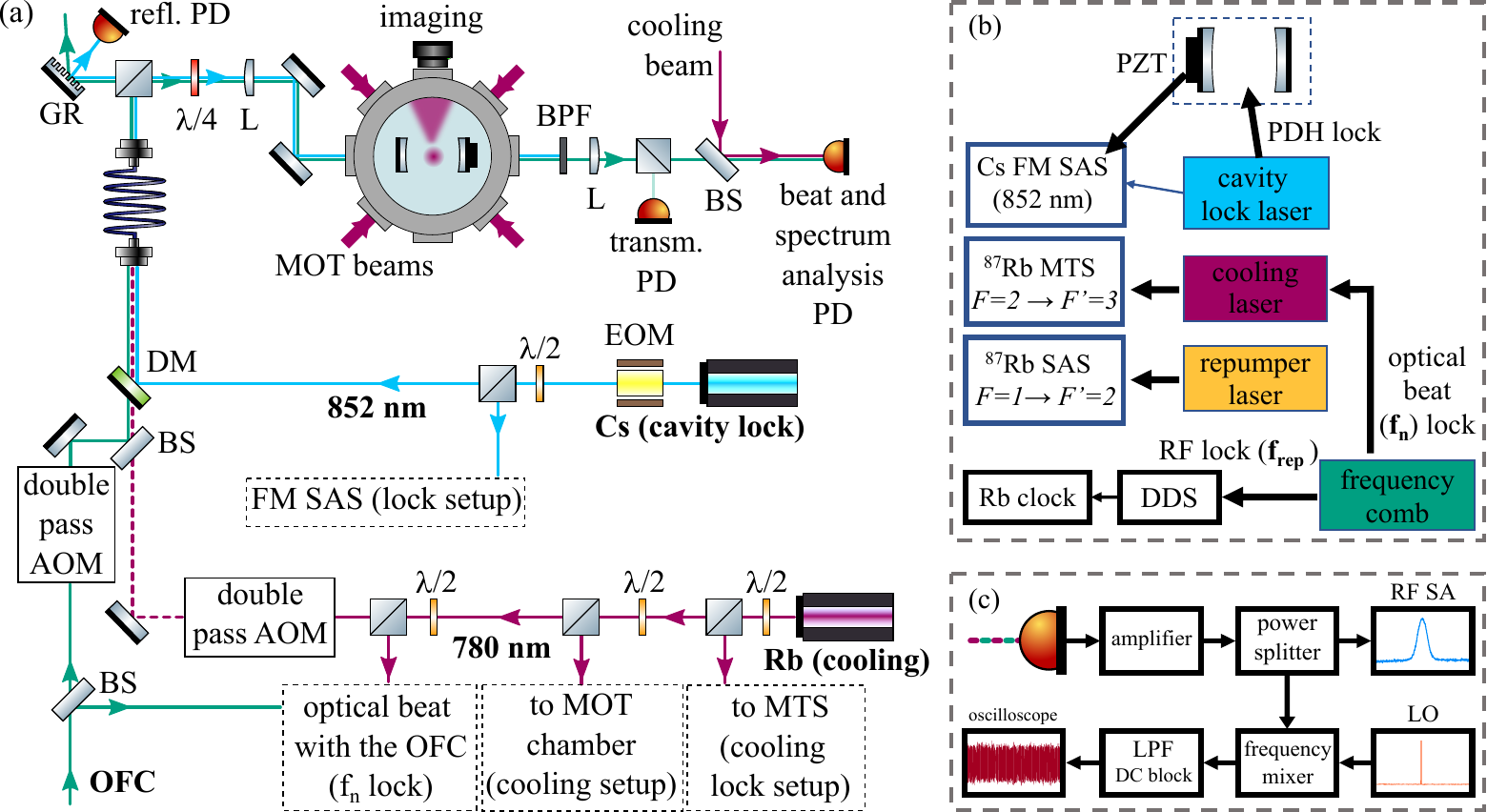}
\caption{(a) Experimental setup. A cloud of cold $^{87}Rb$ atoms is loaded into MOT from background vapor in a stainless-steel vacuum chamber. The center of the MOT is spatially overlapped with the waist of the optical cavity. The cavity is driven by the OFC, and the transmitted light is detected using a photodiode, an optical spectrum analyzer (OSA), and via heterodyne spectroscopy.
$\lambda/2$ – half wave plate, $\lambda$/4 – quarter wave plate, BS – beam splitter, DM – dichroic mirror, GR – grating, L – lens.
(b) Locking scheme. The 852 nm laser is locked to the optical cavity using the Pound–Drever–Hall (PDH) technique, while the cavity itself is stabilized to the 852 nm laser via frequency modulated saturation absorption spectroscopy (FM SAS) in warm cesium vapor.
Both the cooling laser and the repumper laser use a warm rubidium vapor cell in their locking setup. The cooling laser is stabilized to the cooling transition via modulation transfer spectroscopy (MTS), while the repumper laser is stabilized using SAS.
The OFC is stabilized by locking two degrees of freedom: $f_n$ is stabilized to the cooling laser, and $f_{rep}$ is stabilized to a cesium frequency standard-referenced low-noise synthesizer (DDS).
(c) Heterodyne measurement scheme. After amplification, the photodiode signal is split into a monitoring branch (RF spectrum analyzer) and a measurement branch. The measurement branch signal is mixed with a signal generator output and then analyzed using an oscilloscope. LPF – low pass filter, SA – spectrum analyzer, LO – local oscillator.}
\label{SupFig:1}
\end{figure*}

A simplified experimental setup is shown in Fig. \ref{SupFig:1}(a) and the overview of the locking scheme in Fig. \ref{SupFig:1}(b). A cold $^{87}$Rb cloud is loaded from a background vapor in a stainless steel vacuum chamber. The pressure level in the chamber is around 10$^{-8}$ mbar. 
The MOT is realized with the standard six-beam configuration, using a cooling laser (Moglabs MSA003) detuned by $-2\Gamma$ from the $^{87}$Rb  $|5S_{1/2}; F = 2\rangle \rightarrow |5P_{3/2};
F'=3\rangle$ hyperfine transition, and a repumper laser (Toptica DL100) on resonance with the $|5S_{1/2}; F = 1\rangle \rightarrow |5P_{3/2}; F'=2\rangle$ transition. 
The powers of the cooling and repumper lasers, prior to splitting into MOT beams, are 20 mW and 3mW, respectively.
Both beams have a diameter of 1 cm.
The cooling laser is frequency stabilized to the cooling transition using modulation transfer spectroscopy (MTS) and detuned from the resonance using an acousto-optic modulator (AOM) in double-pass configuration. 
The repumper laser is stabilized using saturation absorption spectroscopy (SAS) by modulating the laser current.
This MOT configuration generates a cloud of $\approx7 \times 10^6$ atoms at a temperature of around 70 $\mu$K and a radius of $\approx 0.55$ mm. 

The optical frequency comb (OFC) is generated by frequency doubling an Er:fiber mode-locked femtosecond laser (TOPTICA FFS) operating at 1560 nm with a nominal repetition rate of 80.5 MHz. 
The frequency-doubled spectrum is centered around 780 nm with an FWHM of about 5 nm and a total output power of 38 mW.
To lock the OFC, two degrees of freedom need to be stabilized, for which we choose the laser repetition frequency $f_{rep}$ and the optical frequency of the nth comb mode, $f_n$. 
The repetition frequency $f_{rep} \approx 80.5$ MHz is detected with a high-speed photodiode and referenced to a cesium frequency standard referenced low-noise synthesizer. 
The obtained error signal is used to actively stabilize $f_{rep}$ by feedback to the Er:fiber laser intracavity piezoelectric-transducer-mounted mirror.
The optical frequency $f_n$ is stabilized using heterodyne spectroscopy to an external cavity diode laser which is stabilized to the $^{87}$Rb cooling transition. 
The beat signal between the external laser and the nearest comb mode, $f_{beat}$, is measured and referenced to a cesium frequency standard referenced low-noise synthesizer. 
The obtained error signal is used for stabilization via feedback to the Er:fiber laser current.  
The measured upper limit on the stability of our FC is $\approx$ 3 kHz for an integration time of 10 s.
For details on OFC stabilization, see \cite{santic19}.

The optical cavity consists of two concave mirrors (Layertec 103953) with a radius of curvature R = 7.5 cm and a separation of L = 7.757 cm, corresponding to a near-confocal geometry. 
This configuration results in free spectral range (FSR) of 1.932 GHz, which matches exactly $24 \times f_{rep}$. 
The Gaussian mode waist is calculated to be $w_0 = 100$ $\mu m$, and the single-photon Rabi frequency is $g = 2\pi \times 138$ kHz.   
The cavity mode linewidth, measured via cavity ring-down spectroscopy, is $\kappa = 2\pi \times 150$ kHz, corresponding to a finesse of approximately 12000. 
The cavity length can be tuned by up to 3 $\mu m$ using a piezoelectric transducer (PZT). Detailed information on the cavity design and its characterization is available in \cite{kruljac22}.

The continuous-wave laser (Moglabs CEL cateye laser) at 852 nm is used to stabilize the cavity length. 
This wavelength is chosen because it allows for stabilization of the cavity length using the Cs D2 transition, while being sufficiently far detuned from the $^{87}$Rb resonance to avoid any effect on the dynamics of the cold atoms inside the cavity.
After modulation using an EOM operating at 9.312 MHz, the laser beam is split into two paths. 
One path is used for the stabilization of the 852 nm laser to the cavity via the Pound–Drever–Hall (PDH) technique. 
The other path is directed to a cesium vapor cell, where an error signal is generated through frequency modulated (FM) SAS spectroscopy.
We first lock the laser to the cavity, by using the PDH error signa, via the feedback to the laser current and PZT. 
This approach allows us to lock the laser and cavity together; however, they still experience a common frequency drift over time.
We then use the FM-SAS error signal to stabilize the cavity length via feedback to the cavity PZT.
This closed-loop scheme stabilizes one cavity mode to the Cs atomic transition, thereby locking all cavity modes across a broad frequency range. 
The stability of the cavity length, and consequently the frequency stability of each cavity mode, is limited by the FM-SAS spectroscopy used to generate the error signal, as well as by the performance of the locking electronics. 
It can be estimated to be better than 300 kHz over a 1-second integration time, which corresponds to the measured linewidth of the cavity mode in the absence of atoms.

The OFC light leaking from the cavity is analyzed using different techniques.
A fraction of the transmitted OFC light is monitored using an avalanche photodiode (Thorlabs APD430A/M) and recorded on an oscilloscope. This channel measures the total transmission of all comb modes through the cavity resonances and is used to optimize the matching between the comb and cavity modes.

Spectrally resolved transmission of the OFC is measured using an optical spectrum analyzer (Thorlabs OSA201C) with a resolution of 7.5 GHz. Although this resolution is insufficient to resolve individual comb modes transmitted through the cavity resonances, separated by 1.932 GHz, the measurement still provides valuable information on the total number of comb modes simultaneously transmitted through the cavity. 
The spectra obtained for the purposes of this paper were taken at high resolution and medium-high sensitivity and constructed by averaging 20 traces.

To achieve the single mode coupling configuration (see Fig. 4 of the main text), the OFC comb modes were first shifted in frequency by $u_{m}$ via loading the MOT, after which $\Delta f_0 $ was scanned around the new cavity resonance. Since $u_m$ for $m = \pm 1, \pm 2 $ cavity modes exceeds $\kappa/2\pi $, shifting the OFC by $ u_m $ detunes all other comb modes from their corresponding cavity resonances, i.e. only a single comb mode is resonant with a cavity mode at a time. This single-comb coupling condition was experimentally verified by measuring the cavity transmission spectrum using an OSA, where the transmitted spectrum consisted of a narrow spectral feature limited by the resolution of the OSA. To precisely determine the shape of a cavity resonance, it was therefore necessary to implement a heterodyne detection technique, similar to that in \cite{kresic19}, and described in detail below, which enabled measurements of the transmitted electric field for a single comb line at different values of $\Delta f_0 $.

\begin{figure}[!h]
\centering
\includegraphics[clip,width=\columnwidth]{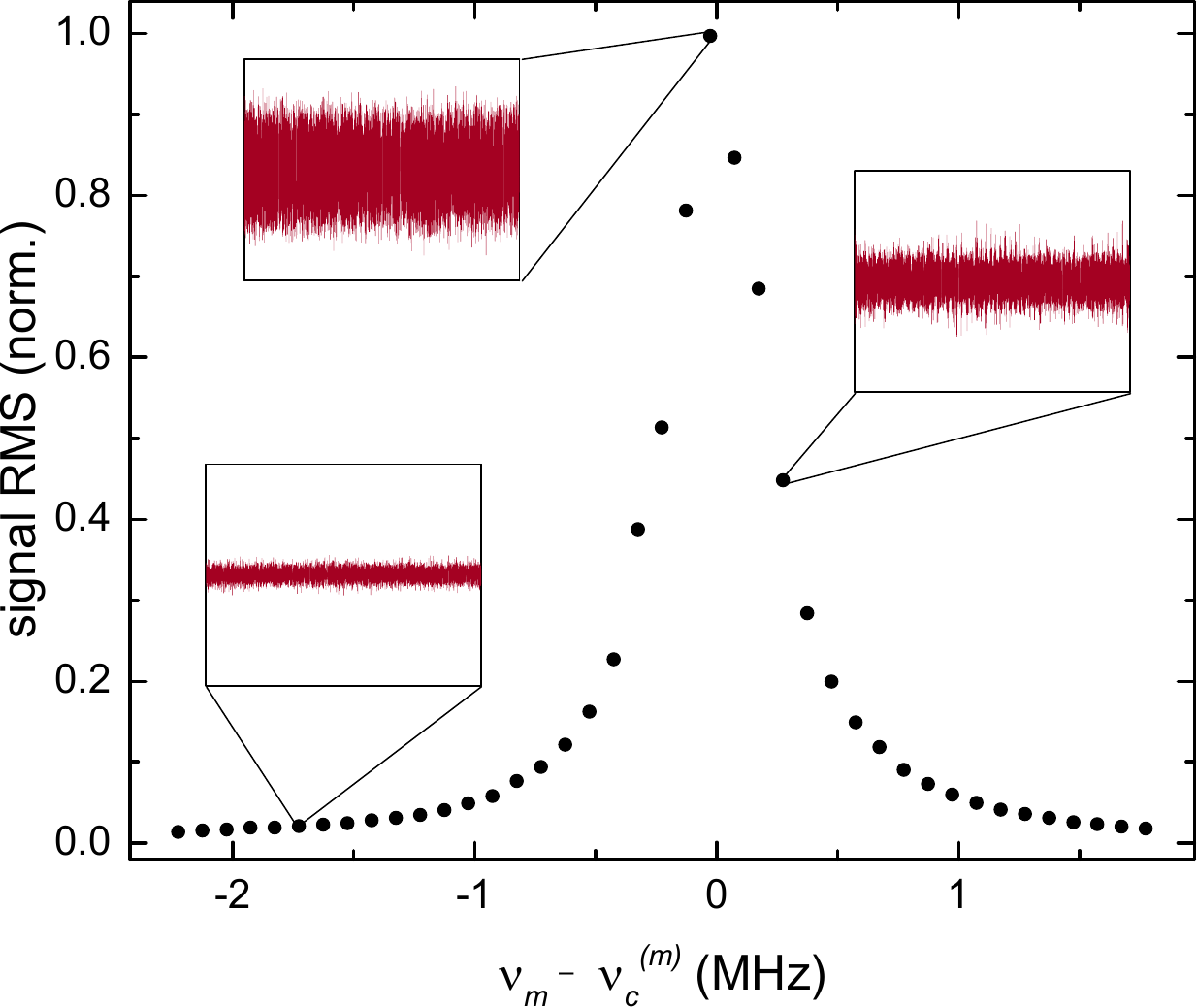}
\caption{RMS beat signals for a transmitted single comb line and an external cw beam, scanned over an empty cavity resonance. The values are proportional to transmitted electric field amplitudes of the single comb line (see text). Each data point represents the RMS of 500000 samples measured over 1 s at the corresponding frequency. The raw data for selected points are given in insets.}
\label{SupFig:2}
\end{figure}

The heterodyne spectroscopy method is employed to resolve and measure light from individual transmitted comb modes.
A fraction of the cooling laser light is combined with the transmitted OFC at a beam splitter and delivered to a fast photodiode (EOT ET-4000) to detect the beat signal.
The signal from the photodiode is amplified by two RF amplifiers (RF-Bay LNA-2535) and split into two paths (see Fig. \ref{SupFig:1}(c)). 
One path, the monitoring branch, is sent to an RF spectrum analyzer to monitor and determine the beat frequency. 
The other path, the measurement branch, is used to measure the amplitude of the beat signal, which is proportional to the electric field of the transmitted comb mode.
In the measurement branch, the RF beat signal is mixed with a tunable local oscillator (Rigol DSG836), which is set to the beat frequency. 
This process demodulates the beat signal. 
After passing through a 240 kHz low-pass filter, the output is proportional to the product of the electric fields of the beat signal and the local oscillator, modulated by the cosine of their phase difference. 
Since the OFC, the cooling laser and the local oscillator are not phase-locked, the resulting signal fluctuates randomly between its minimum and maximum values.
We quantify this fluctuating signal by calculating its root mean square (RMS) value, which is proportional to the electric field amplitude of the transmitted comb mode at the corresponding frequency.
To measure the spectrum of a single cavity mode (see Fig.~\ref{SupFig:2}), we scan the optical frequency of an individual comb mode across the cavity resonance using an AOM, i.e. we scan $\nu_m - \nu_{c}^{(m)}$ where the $\nu_m$ denotes the frequency of the $m$-th empty comb mode and $\nu_{c}^m$ the frequency of the corresponding cavity mode, by scanning $\Delta f_0$.
Simultaneously, we scan the frequency of the local oscillator.
For each value of $\Delta f_0$, we record the fluctuating signal by acquiring 500000 samples over a 1-second interval and calculate the corresponding RMS value.

\begin{figure*}[h!]
\centering
\includegraphics[clip,width=2\columnwidth]{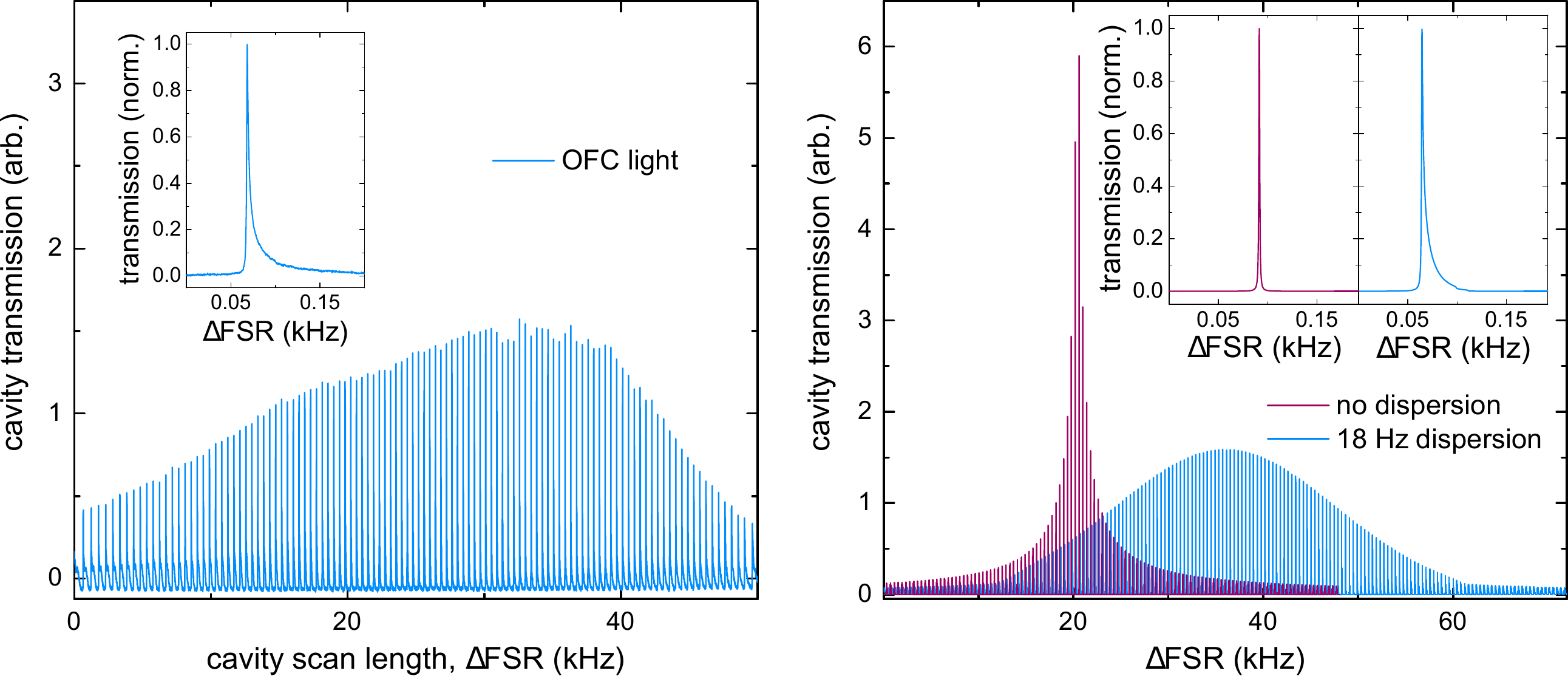}
\caption{a) OFC spectrum transmitted through the cavity, obtained by scanning the full length of the PZT. By looking at an individual peak (inset), we see a "tail", which is a consequence of the cavity mirror dispersion. b) Theoretical calculation of the OFC-cavity transmission with (blue line) and without (burgundy line) dispersion included. Including mirror dispersion results in a broadening and reduction of the transmission.
Inset shows a zoomed-in picture of single peaks for the two cases.}
\label{SupFig:3}
\end{figure*}

\section{OFC-cavity coupling}

The cavity length is designed such that every 24th comb mode can be coupled into the cavity, i.e., $\text{FSR} = 24 \times f_{\text{rep}}$.
To match the optical frequencies of the cavity with the relevant comb modes, we scan the cavity length using a PZT, thereby scanning the FSR, while simultaneously varying $f_{\text{rep}}$.
The OFC-cavity matching condition is achieved when cavity transmission, measured with a photodiode, becomes visible on the oscilloscope.

In Fig.~\ref{SupFig:3}(a) we show measured cavity transmission as a function of the FSR, recorded at a fixed $f_{\text{rep}}$. 
Transmission peaks occur every time the cavity modes in the optical domain shift by $f_{\text{rep}} = 80.532~\text{MHz}$, since then the m-th cavity mode matches the next OFC mode. 
This condition is met when the FSR changes by $\approx$ 0.4~kHz, which corresponds to a cavity length change of  $\approx$ 0.016~\textmu m.
The envelope of the transmitted spectra reflects the coupling efficiency between the OFC and the cavity under different matching conditions.
Maximum transmission peak amplitudes are observed when the cavity FSR is precisely matched to the OFC, such that every 24th comb mode is resonant with a cavity mode, i.e. for the $\Delta f_0=0$. 
For FSR values that are larger or smaller than the optimal FSR values, coupling efficiency falls and the transmission signal is broadened since OFC modes sit in the wings of the cavity resonances, i.e. $\Delta f_0 \neq 0$.

The inset in Fig.~\ref{SupFig:3}(a) shows a zoomed-in view of a single transmission peak, specifically the one corresponding to maximum transmission, where the optimal OFC-cavity matching is achieved.
The transmission peak exhibits an asymmetric shape with a tail that extends toward higher frequencies.
The shape of the individual transmission peak, as well as the broad envelope observed during FSR scanning, arises from intra-cavity dispersion, which introduces a frequency-dependent phase shift $\phi(\omega)$ and consequently makes the FSR frequency dependent.
This reduces the efficiency of the OFC-cavity coupling, as the dispersion term $\partial\phi/\partial\omega$ causes the cavity modes to gradually walk-off from the OFC modes with increasing frequency.

It is important to note that in our experiment we measured the cavity dispersion ($\epsilon$), defined as the difference between two neighboring FSR values, rather than the second-order dispersion term of the mirror reflection phase ($\phi_2$) typically used to specify mirror dispersion in the literature. Using the relation $\epsilon = -4 \pi\, \mathrm{FSR}^3 \, \phi_2 $ \cite{kruljac22}, the measured $\epsilon = 18~\mathrm{Hz}$ corresponds to $\phi_2 = -199~\mathrm{fs}^2$, in great agreement with the value reported by the mirror manufacturer Layertec for 780\,nm ($\phi_2 = -200~\mathrm{fs}^2$). Additional details on the cavity dispersion measurement in our experiment can be found in \cite{kruljac22}.

Figure~\ref{SupFig:3}(b) shows a comparison between the calculated transmission under experimental conditions, both without dispersion and with a dispersion value of 18 Hz. 
Incorporating dispersion into the model yields results that closely match the experimental data.

\begin{figure}[!h]
\centering
\includegraphics[clip,width=\columnwidth]{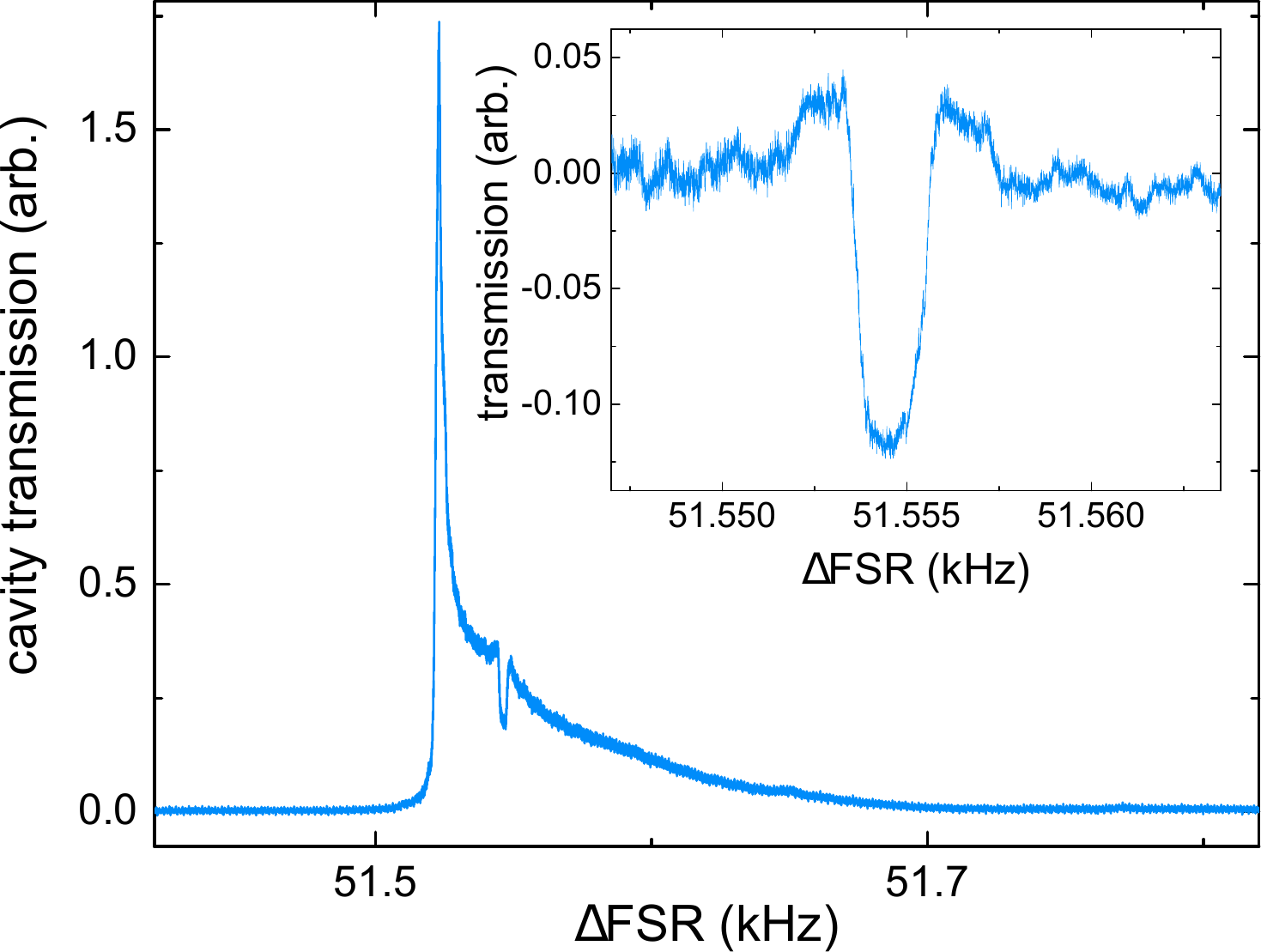}
\caption{A single transmission peak with cold atoms present in the cavity waist. A noticeable dip is visible in the peak's "tail", which signifies an interaction with the atoms. On closer inspection of the dip (inset), one observes both an increase in transmission when the atomic light shift brings the cavity modes closer to the comb modes, and a decrease when it shifts them further apart. This provides the initial evidence of the dispersive atom–cavity interaction.}
\label{SupFig:4}
\end{figure}

We begin the experiment by measuring the transmission of the OFC through an empty cavity. First, the repetition rate$f_{\text{rep}}$ is adjusted to reproduce the transmission spectrum shown in Fig.~\ref{SupFig:3}(a). The cavity FSR is then scanned within a narrow interval around the transmission peak with the highest amplitude.
Next, atoms are loaded into the MOT while continuously monitoring this peak. In the presence of atoms, a pronounced dip appears within the transmission peak (see Fig.~\ref{SupFig:4}), arising from dispersive atom–cavity interactions. 
The number of atoms within the cavity mode is then optimized by maximizing the depth of this transmission dip.
Subsequently, both the cavity length and the OFC are locked to a chosen detuning between the comb mode and the corresponding cavity resonance.
At this stage, we are ready to record the OFC transmission spectra using an OSA or to measure the transmission of a single comb mode via heterodyne spectroscopy.

\section{Cavity quantum electrodynamics model}
\subsection{System Hamiltonian}
We start this section with the Tavis-Cummings Hamiltonian for $N$ two-level atoms interacting with $M$ modes of a Fabry-Perot cavity pumped by a femtosecond comb laser with $M$ modes ($\hbar=1$):
\begin{widetext}
\begin{align}\label{eq:h1at__OFC}
\begin{aligned}
\hat{\mathcal{H}} &= \omega_a\sum_{n=1}^N\hat{\sigma}_{ee}^{(n)}+\left[\sum_{m}\sum_{n=1}^Ng_{m,n}\hat{a}_m\hat{\sigma}_{eg}^{(n)}\cos(k_mx_n)+\mbox{H.c.}\right]+\sum_{m}\omega_{c,m}\hat{a}_{m}^\dagger\hat{a}_{m}+i\sum_{m}\eta_m(\hat{a}_m^\dagger e^{-i\omega_{p,m} t}-\hat{a}_m e^{i\omega_{p,m} t}),
\end{aligned}
\end{align}
\end{widetext}
where $\omega_a$, $\omega_{c,m}$, $\omega_{p,m}$ are the frequencies of the atomic transition, $m$-th cavity mode and $m$-th comb mode, respectively, $g_{m,n}$ is the atom-cavity coupling strength of the $m$-th cavity mode for the $n$-th atom, $\hat{\sigma}_{kk'}^{(n)}=|k\rangle_n\langle k'|_n$ is the atomic optical transition operator of the $n$-th atom, $\eta_m\in\mathbb{R}$ are the pump rates of each comb mode. Note that the kinetic energy is here neglected, such that the studied effects are related only to the electronic degrees of freedom of the two-level atoms. Also, we approximate that the atoms are sufficiently cold for the Doppler shifts to be negligible.

We put $x_n=0$ in Eq. (\ref{eq:h1at__OFC}) for all atoms, such that $\cos(k_m x_n)=1$ and $g_{m,n}\equiv g_0$ for all $m,n$, which would correspond to placing all atoms at the waist of all the cavity modes. By doing this we neglect the dependence of the spatial variation of cavity mode fields on the cavity-atom interaction. This assumption is justified in a cold thermal cloud as the atoms moving by one $\lambda$ at $\sim 1$ ms timescales traverse many wavelengths in the $\gtrsim 1$s timescales of the experiment.

We also write the collective Pauli operators as $\hat{\sigma}_{kk'}=\sum_{n=1}^N\hat{\sigma}_{kk'}^{(n)}$. The above stated approximations lead to a Hamiltonian of the form:
\begin{widetext}
\begin{align}\label{eq:h1at_2}
\begin{aligned}
\hat{\mathcal{H}} &= \omega_a\hat{\sigma}_{ee}+g_0\left[\sum_{m} \hat{a}_m\hat{\sigma}_{eg}+\mbox{H.c.}\right]+\sum_{m}\omega_{c,m}\hat{a}_{m}^\dagger\hat{a}_{m}+i\eta\sum_{m}(\hat{a}_m^\dagger e^{-i\omega_{p,m} t}-\hat{a}_m e^{i\omega_{p,m} t}),
\end{aligned}
\end{align}
\end{widetext}
The Hamiltonian (\ref{eq:h1at_2}) can be converted into the form:
\begin{widetext}
\begin{align}\label{eq:h1at_3}
\begin{aligned}
\hat{H}(t) &=\hat{U}\hat{\mathcal{H}}\hat{U}^\dagger+i(\partial_t\hat{U})\hat{U}^\dagger=-\sum_{m}\Delta_{c,m}\hat{a}_{m}^\dagger\hat{a}_{m}+g_{0}\left[\sum_{m}\hat{a}_me^{-i\delta_{p,m}t}\hat{\sigma}_{eg}+\mbox{H.c.}\right]+i\sum_{m}\eta_m(\hat{a}_m^\dagger -\hat{a}_m ),
\end{aligned}
\end{align}
\end{widetext}
where $\delta_{p,m}=\omega_{p,m}-\omega_a$, $\Delta_{c,m}=\omega_{p,m}-\omega_{c,m}$, by using
\begin{align}\label{eq:U_op}
\begin{aligned}
\hat{U} &=\exp\left[i t\left(\omega_a\hat{\sigma}_{ee}+\sum_{m}\omega_{p,m}\hat{a}_{m}^\dagger\hat{a}_{m}\right)\right].
\end{aligned}
\end{align}

\subsection{Schrieffer-Wolff transformation}
The Hamiltonian (\ref{eq:h1at_3}) can be written as $\hat{H}(t)=\hat{H}_0+\hat{H}_{drive}+\hat{V}(t)$, where 
\begin{align}\label{eq:H_decompose}
\begin{aligned}
\hat{H}_0 &= -\sum_{m}\Delta_{c,m}\hat{a}_{m}^\dagger\hat{a}_{m},\:
\hat{H}_{drive} = i\sum_{m}\eta_m(\hat{a}_m^\dagger -\hat{a}_m ),\\
\hat{V}(t) & = g_0\left[\sum_{m}\hat{a}_me^{-i\delta_{p,m}t}\hat{\sigma}_{eg}+\mbox{H.c.}\right],
\end{aligned}
\end{align}
and $\hat{V}(t)$ is considered a small perturbation. The Schrieffer-Wolff transformation \cite{schrieffer66} expresses the Hamiltonian $\hat{H}(t)$ in the ``dressed state" basis by performing a rotation $\hat{H}'(t)=\hat{U}_{S}\hat{H}(t)\hat{U}_{S}^\dagger+i(\partial_t\hat{U}_S)\hat{U}_{S}^\dagger$ with $\hat{U}_{S}(t)=e^{\hat{S}(t)}$, where the anti-Hermitian operator $\hat{S}(t)$ is the generator of the unitary transformation $\hat{U}_{S}$. The first term in the Hamiltonian $\hat{H}'(t)$ can be written using the Baker-Hausdorff expansion, while for calculating the second term one first uses Eq. (2.1) of Ref. \cite{wilcox_exponential_1967} to calculate
\begin{align}\label{eq:calc_Uterm}
\begin{aligned}
(\partial_t\hat{U}_S)\hat{U}_S^\dagger &= \int_0^1e^{u\hat{S}(t)}(\partial_t \hat{S}(t))e^{-u\hat{S}(t)}du,
\end{aligned}
\end{align}
which can be expanded using the Baker-Hausdorff lemma and after integrating over the $u$ terms one has:
\begin{align}\label{eq:calc_Uterm}
\begin{aligned}
(\partial_t\hat{U}_S)\hat{U}_S^\dagger &= \partial_t\hat{S}+\frac{1}{2}[\hat{S},\partial_t\hat{S}]+\frac{1}{6}[\hat{S},[\hat{S},\partial_t\hat{S}]]+...
\end{aligned}
\end{align}
The expansion for $\hat{H}'(t)$ now has the form \cite{wang24}:
\begin{align}\label{eq:ham_prime}
\begin{aligned}
\hat{H}'(t)&=\hat{H}+[\hat{S},\hat{H}]+\frac{1}{2}[\hat{S},[\hat{S},\hat{H}]]+i\partial_t\hat{S}+\frac{i}{2}[\hat{S},\partial_t\hat{S}]+...
\end{aligned}
\end{align}
Taking the $\hat{S}$ which satisfies:
\begin{align}\label{eq:s_cond}
\begin{aligned}
\hat{V}&+[\hat{S},\hat{H}_0]+i\partial_t\hat{S}=0,
\end{aligned}
\end{align}
one can remove the terms first order in $g_0$ and get the Hamiltonian:
\begin{align}\label{eq:ham_prime1}
\begin{aligned}
\hat{H}'&=\hat{H}_0+\hat{H}_{drive}+[\hat{S},\hat{H}_{drive}]+\frac{1}{2}[\hat{S},\hat{V}]+...
\end{aligned}
\end{align}
The operator $\hat{S}$ satisfying condition (\ref{eq:s_cond}) is given by:
\begin{align}\label{eq:S_form}
\hat S(t) \;=\;- g_0 \sum_{m}
\frac{\hat a_m^\dagger e^{i\delta_{p,m}t}\,\hat\sigma_{ge}-\hat a_m e^{-i\delta_{p,m}t}\,\hat\sigma_{eg} }
     {\Delta_{c,m}-\delta_{p,m}} .
\end{align}
\newpage
Defining now $\Delta_a^{(m)}=\omega_{c,m}-\omega_a$ as in the main text, we have
\begin{align}\label{eq:S_form_final}
\hat S(t) \;=\; g_0 \sum_{m}
\frac{\hat a_m^\dagger e^{i\delta_{p,m}t}\,\hat\sigma_{ge}-\hat a_m e^{-i\delta_{p,m}t}\,\hat\sigma_{eg} }
     {\Delta_a^{(m)}} .
\end{align}

\subsection{Effective Hamiltonian}
Plugging in the expression (\ref{eq:S_form_final}) into (\ref{eq:ham_prime1}), the Hamiltonian $\hat{H}'(t)$ up to the second order in perturbation [where the neglected terms are $\sim O\left(g_0^3/|\Delta_{a}^{(m)}|^2,\:g_0^2\eta_m/|\Delta_{a}^{(m)}|^2\right) $] is given by:
\begin{widetext}
\begin{align}\label{eq:Ham_prime2}
\begin{aligned}
\hat H'(t) &= -\sum_m \Delta_{c,m}\,\hat a_m^\dagger\hat a_m
   + i\sum_m \eta_m(\hat a_m^\dagger-\hat a_m) -\sum_m \frac{g_0^2}{\Delta_{a}^{(m)}}\,\left(\hat a_m^\dagger\hat a_m+\frac{1}{2}\right)\hat\sigma_z-\frac{1}{2}
   \sum_m \frac{g_0^2}{\Delta_{a}^{(m)}}\sum_{i=1}^N\sum_{j\neq i}^N\left(\hat{\sigma}_{eg}^{(i)}\hat{\sigma}_{ge}^{(j)}+\mbox{h.c.}\right) \\
   -&\frac{1}{2}\sum_m\sum_{n\neq m}\frac{g_0^2}{\Delta_{a}^{(m)}}
       \Big( e^{i(\omega_{p,n}-\omega_{p,m})t}\,\hat a_n^\dagger\hat a_m + \mathrm{h.c.} \Big)\hat\sigma_z + \sum_m \left(\frac{ig_0\eta_m}{\Delta_{a}^{(m)}}e^{i\delta_{p,m}t}\hat\sigma_{ge}+\mbox{h.c.}\right).
\end{aligned}
\end{align}
\end{widetext}
The first two terms are $\hat{H}_0$ and $\hat{H}_{drive}$. The third term gives the light shift of the cavity modes due to coupling to the atoms, and the light shifts of the atomic optical transition due to coupling to the driven cavity. The fourth term describes cavity-mediated dipole-dipole interaction between the atoms. This term vanishes in the mean field limit and can otherwise be neglected if one assumes that the thermal atomic motion during the experiment is on the scales larger than the cavity wavelengths. Note that the approximation of modeling the cold cloud as an effective homogeneus medium while neglecting the kinetic energy and dipole-dipole scattering between the atoms is commonly used in deriving the mean-field semiclassical Maxwell-Bloch type equations for atomic dipoles in a single mode cavity \cite{arecchi65,gabor23}. The assumptions should also hold in the many longitudinal mode case considered here since the wavenumbers $k$ of the modes do not vary appreciably on the scales of the transmitted OFC beam, as the wavelength spread is $\Delta \lambda\sim 1$ nm.

The fifth term introduces scattering of the cavity photons from one mode into another, mediated by the atoms, demonstrated for the multimode degenerate transverse mode case in e.g. \cite{wickenbrock_collective_2013}. The sum terms in our case of longitudinal cavity modes are fast-oscillating with respect to the interaction strength, such that the condition $\frac{Ng_0^2\langle\hat{a}_m^\dagger\hat{a}_m\rangle}{|\Delta_{a}^{(m)}|}\ll|\omega_{p,n}-\omega_{p,m}|$ is satisfied in the dispersive linear limit (see also below) and they can be neglected. The last term represents the driving of the atomic transition by the cavity photons pumped by the OFC pulses with repetition frequency given by the cavity FSR. 

Neglecting the dipole-dipole and cavity cross-coupling terms, we arrive at the effective Hamiltonian of our system:
\begin{widetext}
\begin{align}\label{eq:H_eff}
\begin{aligned}
\hat H_{eff}(t)&= -\sum_m \Delta_{c,m}\,\hat a_m^\dagger\hat a_m
   + i\sum_m \eta_m(\hat a_m^\dagger-\hat a_m) -\sum_m \frac{g_0^2}{\Delta_{a}^{(m)}}\,\left(\hat a_m^\dagger\hat a_m+\frac{1}{2}\right)\hat\sigma_z+ \sum_m \left(\frac{ig_0\eta_m}{\Delta_{a}^{(m)}}e^{i\delta_{p,m}t}\hat\sigma_{ge}+\mbox{h.c.}\right).
\end{aligned}
\end{align}
\end{widetext}
Comparing the maximal interaction strength in the first order perturbed Hamiltonian $\sim \frac{Ng_0^2\langle\hat{a}_1^\dagger\hat{a}_1\rangle}{\Delta_a^{(1)}}$, with the $\hat{V}$'s minimal oscillation frequency $\sim \Delta_a^{(1)}$ (equivalent to the minimum energy splitting of the unperturbed Hamiltonian in the static case \cite{blais21}), we get the validity of the perturbative expansion (\ref{eq:H_eff}) to be $\frac{g_0\sqrt{N\langle\hat{a}_1^\dagger\hat{a}_1\rangle}}{|\Delta_{a}^{(1)}|}\ll 1$. This is satisfied in the linear optics, far detuned (dispersive) atom-cavity limit used in our experiment, as confirmed ultimately by comparison to the measured results.

\subsection{Cavity transmission}
We now consider the steady state intracavity photon population $\langle\hat{a}_m^\dagger\hat{a}_m\rangle$ in the thermodynamic mean field limit. The Lindblad equation for $\hat{\rho}_{eff}(t)=\hat{U}_S(t)\hat{U}(t)\hat{\rho}(t)\hat{U}^\dagger(t)\hat{U}_S^\dagger(t)$ is now:
\begin{align}\label{eq:lindblad_sw}
\begin{aligned}
\frac{d\hat{\rho}_{eff}}{dt} =&-i[\hat{H}_{eff}(t),\hat{\rho}_{eff}]+\kappa\sum_m\left(\hat{a}'_m\hat{\rho}_{eff}\hat{a}_m'^\dagger-\frac{1}{2}\{\hat{a}_m'^\dagger\hat{a}_m',\hat{\rho}_{eff} \}\right)\\
+&\Gamma\sum_{n=1}^N\left(\hat{\sigma}_{ge}'^{(n)}\hat{\rho}_{eff}\hat{\sigma}_{eg}'^{(n)}-\frac{1}{2}\{\hat{\sigma}_{eg}'^{(n)}\hat{\sigma}_{ge}'^{(n)},\hat{\rho}_{eff} \}\right),
\end{aligned}
\end{align}
where $\kappa$, $\Gamma$ are cavity and atomic decay rates, $\hat{a}_m'=\hat{U}_S(t)\hat{U}(t)\hat{a}_m\hat{U}^\dagger(t)\hat{U}_S^\dagger(t)=e^{-i\omega_{p,m}t}\hat{U}_S(t)\hat{a}_m\hat{U}_S^\dagger(t)$, $\hat{\sigma}_{eg}'^{(n)}=\hat{U}_S(t)\hat{U}(t)\hat{\sigma}_{eg}^{(n)}\hat{U}^\dagger(t)\hat{U}_S^\dagger(t)=e^{i\omega_{a}t}\hat{U}_S(t)\hat{\sigma}_{eg}^{(n)}\hat{U}_S^\dagger(t)$, and we have neglected the coherence in the spontaneous decay of atomic dipoles (see also above). Using the Baker-Hausdorff lemma, the decay operators can be calculated to give:
\begin{align}\label{eq:transformed_decay}
\begin{aligned}
\hat{U}_S(t)\hat{a}_m\hat{U}_S^\dagger(t)& =\hat{a}_m -\frac{g_0}{\Delta_a^{(m)}}e^{i\delta_{p,m}t}\hat{\sigma}_{ge}+...,\\
\hat{U}_S(t)\hat{\sigma}_{eg}^{(n)}\hat{U}_S^\dagger(t)& =\hat{\sigma}_{eg}^{(n)}-g_0\hat{\sigma}_{z}^{(n)}\sum_m\frac{1}{\Delta_a^{(m)}}\hat{a}_m^\dagger e^{i\delta_{p,m}t}+...
\end{aligned}
\end{align}
We can now use the equation for the evolution of expectation value of an operator $\langle\hat{O}\rangle=\mbox{Tr}[\hat{O}\hat{\rho}_{eff}(t)]$:
\begin{align}\label{eq:opexpect_sw}
\begin{aligned}
\frac{d\langle \hat{O}\rangle }{dt}&=i\langle [\hat{H}_{eff}(t),\hat{O}]\rangle +\kappa\sum_m\left\langle \hat{a}_m'^\dagger \hat{O} \hat{a}_m'-\frac{1}{2}\{\hat{a}_m'^\dagger\hat{a}_m',\hat{O}\}\right\rangle\\
+&\Gamma\sum_{n=1}^N\left\langle\hat{\sigma}_{eg}'^{(n)}\hat{O}\hat{\sigma}_{ge}'^{(n)}-\frac{1}{2}\{\hat{\sigma}_{eg}'^{(n)}\hat{\sigma}_{ge}'^{(n)},\hat{O} \}\right\rangle.
\end{aligned}
\end{align}
The evolution equation for $\hat{a}_m$ thus has the form:
\begin{align}\label{eq:lindcav}
\begin{aligned}
\frac{d\langle \hat{a}_m\rangle }{dt}&=i\langle [\hat{H}_{eff}(t),\hat{a}_m]\rangle -\frac{\kappa}{2}\langle \hat{a}_m\rangle \\
+&\frac{\Gamma}{2}\left(\frac{g_0}{\Delta_a^{(m)}}e^{i\delta_{p,m}}\langle\hat{\sigma}_{ge}\rangle-\frac{g_0^2}{(\Delta_a^{m})^2}\langle\hat{a}_m\hat{\sigma}_{z}\rangle\right).
\end{aligned}
\end{align}
Note that the third term is fast-oscillating, while it also in the mean field thermodynamic limit has the magnitude $\left|\frac{\Gamma g_0N\sigma_{ge}}{2\Delta_a^{(m)}}\right|\ll\eta_m$, and can thus be neglected with respect to $\eta_m$. The fourth term in the mean field thermodynamic limit has the magnitude $\frac{\Gamma Ng_0^2}{2(\Delta_a^{(m)})^2}\langle\hat{a}_m\rangle\ll\kappa\langle\hat{a}_m\rangle$ and can thus be neglected with respect to the $\kappa\langle\hat{a}_m\rangle$ term. The operator $\hat{a}_m$ can now be approximated to evolve as:
\begin{align}\label{eq:cav_evol}
\begin{aligned}
\frac{d\langle \hat{a}_m\rangle }{dt}&=\left(i\Delta_{c,m}-\frac{\kappa}{2}\right)\langle\hat{a}_m\rangle+i\frac{g_0^2}{\Delta_a^{(m)}}\langle\hat{\sigma}_z\hat{a}_m\rangle+\eta_m.
\end{aligned}
\end{align}
In the mean-field thermodynamic limit $\langle\hat{a}_m\rangle\to\sqrt{N}\alpha_m,\:\langle\hat{\sigma}_z\rangle\to N\sigma_z=N(\sigma_{ee}-\sigma_{gg})$ and $\langle\hat{O}_1\hat{O}_2\rangle\to\langle\hat{O}_1\rangle\langle\hat{O}_2\rangle$. For the linear optics limit, one can neglect the excited state population, i.e. $\sigma_z\approx -1$, such that the steady state cavity mode population $|\alpha_m|^2$ is given by:
\begin{align}\label{eq:final_cavpop}
\begin{aligned}
|\alpha_m|^2=\frac{1}{N}\frac{|\eta_m|^2}{\left(\Delta_{c,m}-\frac{Ng_0^2}{\Delta_a^{(m)}}\right)^2+\frac{\kappa^2}{4}}.
\end{aligned}
\end{align}
In the dispersive limit, the presence of the atoms thus leads to collective cavity light shifts given by $Ng_0^2/\Delta_a^{(m)}$ for the $m$-th cavity mode. Note that the classical calculations for an $m$-th cavity mode follow exactly the same Lorentzian line shape and collective light shift, with cavity parameters rescaled as shown in Eq. (\ref{eq:transm_lorentzian}).

\section{Cavity transmission spectrum - classical model}
The origin of the collective light shift can also be explained using classical optics. At large cavity-atom detunings $|\Delta_a|=|\omega_c-\omega_a|\gg \Gamma_2$ ($\omega_c=2\pi\times\nu_c$ - cavity mode frequency, $\omega_a=2\pi\times\nu_a$ - atomic optical transition frequency, $\Gamma_2$ - optical coherence decay rate), absorption is suppressed by a factor of $\Gamma_2/\Delta_a$ with respect to refraction, and the medium can be approximated as purely dispersive. For a homogeneous medium with effectively $N$ atoms in a volume $V$ inside an idealized plano-planar Fabry-Perot cavity, every mode frequency then shifts to $\omega_c'=\omega_c/n_{a}$, due to change in the optical length of the cavity, where $1-n_{a}\propto N/(V\Delta_a)$. Calculating the steady state refractive index of two-level atoms $n_a$, the effective collective light shift of the cavity mode is then given by $\omega_c'-\omega_c=Ng_0^2/\Delta_a$, where $g_0$ is the cavity-atom coupling strength (see below). This frequency shift is routinely observed for cold and ultracold atoms coupled to a single cavity mode pumped by a cw laser \cite{ritsch13,gothe19}.

In the classical linear optics model, the transmission of every OFC mode through the cavity is viewed separately, i.e. the comb is modeled as a set of cw lasers \cite{thorpe2008cavity}. The frequency of a $q$-th longitudinal mode of an empty cavity is given by $\omega_c=2\pi q/t_{RT}$, and $t_{RT}=2n_{r}L/c_0$ is the round-trip time, where $q$ is an integer, $L$ is the cavity length and $n_{r}\approx 1+\chi_r/2$ the real part of the (generally nonlinear) refractive index of the cavity medium at pump frequency $\omega\approx\omega_c$, which is given by 1 for an empty cavity \cite{ismail16}. If we neglect absorption, the difference between empty cavity frequency $\omega_c$ and the with-atom cavity frequency $\omega_c'$ is thus given by: 
\begin{align}\label{eq:prelim1}
\begin{aligned}
 \omega_c-\omega_c'=\frac{qc_0}{2n_{at}L}(n_{at}-1)\approx \omega_c\frac{\chi_{at}}{2} 
\end{aligned}
\end{align}
where we have (just like in deriving $n=\sqrt{1+\chi}\approx 1+\chi/2$) neglected the absorption and quadratic terms in $\chi$, approximating $(n_{at}-1)/n_{at}\approx \chi_{at}/2$. 


The real part of $\chi$ for two-level atoms with frequency $\omega_a$ pumped by a beam of intensity $I$ and frequency $\omega$ is for the steady state given by \cite{boyd_nonlinear_2008,foot_atomic_2005}:
\begin{align}\label{eq:twolevel_1}
\begin{aligned}
\chi_{at} &= \frac{P_{at}^{NL}}{\varepsilon_0 E}=\frac{N\mu}{V\varepsilon_0 E}\mbox{Re}(\rho_{eg})=\frac{N\mu^2}{V\hbar\varepsilon_0 }\frac{\Delta_a}{\Delta_a^2+\frac{\Gamma^2}{4}}(\rho_{ee}-\rho_{gg})\\
=&-\frac{N\mu^2}{V\hbar\varepsilon_0 }\frac{\Delta_a}{\Delta_a^2+\frac{\Gamma^2}{4}}\frac{\Delta_a^2+\frac{\Gamma^2}{4}}{\Delta_a^2+\frac{\Omega^2}{2}+\frac{\Gamma^2}{4}}
\end{aligned}
\end{align}
where $\Delta_a=\omega-\omega_a$, Rabi frequency $\Omega=\Gamma\sqrt{I/(2I_s)}$, $\Gamma$ - atomic population decay rate. In the limit $\Delta_a\gg \Gamma$, this can be approximated by:
\begin{align}\label{eq:twolevel_2}
\chi_{at} &\approx -\frac{N\mu^2}{V\hbar\varepsilon_0 \Delta_a}\frac{1}{1+\frac{\Gamma^2I}{4\Delta_a^2 I_{s}}}.
\end{align}
For $|\omega-\omega_c|\ll|\Delta_a|$, one can replace $\omega\to\omega_c$ in the above equations, and the linear collective light shift of each cavity mode can then be calculated using Eq. (\ref{eq:prelim1}) to give \cite{gothe19}: 
\begin{align}\label{eq:twolevel_3}
\omega_c-\omega_c' &= -\frac{Ng_0^2}{\Delta_a}\frac{1}{1+\frac{\Gamma^2I}{4\Delta_a^2 I_{s}}}\approx -\frac{Ng_0^2}{\Delta_a}=-2\pi \times u_0,
\end{align}
where we have used $\hbar g_0=\mu\sqrt{\frac{\hbar\omega_c}{2\varepsilon_0 V}}$ \cite{ritsch13}.

The intensity transmission coefficient $T_m(\nu_m)$ of an OFC line at frequency $\nu_m$ can be well approximated by a Lorentzian \cite{ismail16}:
\begin{align}\label{eq:transm_lorentzian}
T_m(\nu_m)=\frac{(1-R)^2}{(1-R)^2+4R^2\phi_m^2(\nu_m)},
\end{align}
with $R=\sqrt{r_1r_2}$ - intensity reflectivity of the cavity mirrors, $\phi_m(\nu_m)=(\nu_m-\nu_{c,m}')/\Delta\nu_{FSR}^m$ and $m$ is an integer between $-N_{mod}/2$ and $N_{mod}/2$ excluding 0, with $N_{mod}$ being the total number of relevant cavity modes. 

Due to mirror dispersion, $m$-th cavity mode shifts from the perfect resonance $\nu_{c,per}^{(m)}$, and the empty cavity resonances are given by $\nu_{c}^{(m)}=\nu_{c,per}^{(m)}+|m|(|m|-1)\epsilon/2$, while the FSR changes with $m$ as $\Delta\nu_{FSR}^m=\Delta\nu_{FSR}^0+m\epsilon$ for positive, and $\Delta\nu_{FSR}^m=\Delta\nu_{FSR}^0+(m+ 1)\epsilon$ for negative $m$ values. Noting that $|m|\epsilon/\Delta\nu_{FSR}^0\ll 1$ for all $m$, we can approximate $\Delta\nu_{FSR}^m\approx \Delta\nu_{FSR}^0$. Using Eq. (\ref{eq:twolevel_3}), we finally have the expression for the $m$-th OFC mode detuning from the corresponding cavity mode: $\nu_m-\nu_{c,m}'=\Delta f_0-|m|(|m|-1)\epsilon/2-u_m$.

\section{Single comb line coupling regime - mean field quantum dynamics}
In this model, the cavity is driven by a laser with frequency $\omega$ while the atoms are driven by another laser with frequency $\omega_M$, where the wavenumbers $k$ are approximately equal for the two beams. The cavity-atom coupling will here be described by the Tavis-Cummings model, while the atomic drive is described by a classical pump field with a Rabi frequency $\Omega_M$. The system Hamiltonian for $N$ two-level atoms in the dipolar and rotating wave approximations is here given by ($\hbar=1$) \cite{tavis68}:
\begin{align}\label{eq:h1at_1}
\begin{aligned}
\hat{\mathcal{H}} &= \omega_a\sum_{n=1}^N\hat{\sigma}_{ee}^{(n)}
+\omega_{c}\hat{a}^\dagger\hat{a}+i\eta(\hat{a}^\dagger e^{-i\omega t}-\hat{a} e^{i\omega t})\\
+&\left[\sum_{n=1}^N(g_{0}\hat{a}\cos(kx_n)+\Omega_Me^{i(kx_n-\omega_M t)})\hat{\sigma}_{eg}^{(n)}+\mbox{H.c.}\right],
\end{aligned}
\end{align}
where $\omega_a$ and $\omega_{c}$ are the frequencies of the atomic transition and cavity mode, $g_{0}$ is the atom-cavity coupling strength and $\hat{\sigma}_{jj'}^{(n)}=|j\rangle_n\langle j'|_n$ are the raising, lowering and population operators of the $n$-th atom, $\eta\in\mathbb{R}$ is the cavity pump rate, while $\hat{a}$ is the annihilation operator for the cavity mode. Note that the kinetic energy is here neglected, such that the studied effects are related only to the electronic degrees of freedom of the two-level atoms. As above, we approximate the atoms to be sufficiently cold for the Doppler shifts to be negligible. As the atoms in the cloud are in thermal motion during the experiment, we put $x_n=0$ in Eq. (\ref{eq:h1at_1}) for all atoms, i.e. we neglect the dependence of the atom-light interaction on atomic position as the atoms traverse many wavelengths during the duration of the experiment (see also above). 

Writing the collective atomic operators as $\hat{\sigma}_{jj'}=\sum_{n=1}^N\hat{\sigma}_{jj'}^{(n)}$ leads to a Hamiltonian of the form:
\begin{align}\label{eq:hmot1}
\begin{aligned}
\hat{\mathcal{H}} &= \omega_a\hat{\sigma}_{ee}+(g_0\hat{a}\hat{\sigma}_{eg}+\Omega_Me^{-i\omega_M t}\hat{\sigma}_{eg}+\mbox{H.c.})\\
+&\omega_{c}\hat{a}^\dagger\hat{a}+i\eta(\hat{a}^\dagger e^{-i\omega t}-\hat{a} e^{i\omega t}).
\end{aligned}
\end{align}
Going into the rotating frame of the cavity pump beam using the operator $\hat{U}=\exp\left[i\omega t\left(\hat{\sigma}_{ee}+\hat{a}^\dagger\hat{a}\right)\right]$ leads to the Lindblad equation for $\hat{\rho}_{rot}=\hat{U}\hat{\rho}\hat{U}^\dagger$:
\begin{align}\label{eq:lindblad1}
\begin{aligned}
\frac{d\hat{\rho}_{rot}}{dt} =&-i[\hat{H}(t),\hat{\rho}_{rot}]+\kappa\left(\hat{a}\hat{\rho}_{rot}\hat{a}^\dagger-\frac{1}{2}\{\hat{a}^\dagger\hat{a},\hat{\rho}_{rot} \}\right)\\
+&\Gamma\left(\hat{\sigma}_{ge}\hat{\rho}_{rot}\hat{\sigma}_{eg}-\frac{1}{2}\{\hat{\sigma}_{eg}\hat{\sigma}_{ge},\hat{\rho}_{rot} \}\right).
\end{aligned}
\end{align}
The Hamiltonian is now given by: 
\begin{align}\label{eq:hmot2}
\begin{aligned}
\hat{H}(t) =&\hat{U}\hat{\mathcal{H}}\hat{U}^\dagger+i(\partial_t\hat{U})\hat{U}^\dagger\\
=-&\Delta_a\hat{\sigma}_{ee}+(g_{0}\hat{a}\hat{\sigma}_{eg}+\Omega_Me^{i\Delta_M t}\hat{\sigma}_{eg}+\mbox{H.c.})\\
-&\Delta_{c}\hat{a}^\dagger\hat{a}+i\eta(\hat{a}^\dagger -\hat{a} ),
\end{aligned}
\end{align}
where $\Delta_a=\omega-\omega_a$, $\Delta_M=\omega-\omega_M$ and $\Delta_c=\omega-\omega_c$. 

The dynamical equations for the atomic operator and cavity mode expectation values $\langle \hat{O}\rangle=\mbox{Tr}(\hat{O}\hat{\rho}_{rot} (t))$ can be derived using the Lindblad equation, which results in:
\begin{align}\label{eq:lind1}
\begin{aligned}
\frac{d\langle \hat{O}\rangle }{dt}&=i\langle [\hat{H}(t),\hat{O}]\rangle +\kappa\langle \hat{a}^\dagger \hat{O} \hat{a}-\frac{1}{2}\{\hat{a}^\dagger\hat{a},\hat{O}\}\rangle\\
+&\Gamma\langle\hat{\sigma}_{eg}\hat{O}\hat{\sigma}_{ge}-\frac{1}{2}\{\hat{\sigma}_{eg}\hat{\sigma}_{ge},\hat{O} \}\rangle.
\end{aligned}
\end{align}
In the thermodynamic mean field limit, where $\langle\hat{O}_1\hat{O}_2\rangle\to\langle\hat{O}_1\rangle\langle\hat{O}_2\rangle$ and $\langle \hat{a}\rangle\to \sqrt{N}\alpha$, $\langle\hat{\sigma}_{ij}\rangle\to N\sigma_{ij}$, using $\langle\hat{\sigma}_{eg}\hat{\sigma}_{ge}\rangle\to \langle\hat{\sigma}_{ee}\rangle$ (i.e. neglecting again the atomic dipole-dipole correlations) and $\langle\hat{\sigma}_{gg}+\hat{\sigma}_{ee}\rangle=N$, the equations of motion are given by 
\begin{align}\label{eq:dynevol4}
\frac{\partial \alpha}{\partial t}&=\left(i\Delta_{c}-\frac{\kappa}{2}\right)\alpha-ig_N\sigma_{ge}+\frac{\eta}{\sqrt{N}},\\
\frac{\partial \sigma_{ee}}{\partial t}&=-\Gamma\sigma_{ee}+\left[i\left(g_N\alpha^*+\Omega_M^*e^{-i\Delta_M t}\right) \sigma_{ge}+\mbox{H.c.}\right] ,\\
\frac{\partial \sigma_{eg}}{\partial t}&=-\left(i\Delta_{a}+\frac{\Gamma}{2}\right)\sigma_{eg}+i\left(g_N\alpha^*+\Omega_M^*e^{-i\Delta_M t}\right) (1-2\sigma_{ee}),
\end{align}
where $g_N=\sqrt{N}g_0$.
